\documentclass[prl,aps,amsmath,amssymb, preprintnumbers, reprint,longbibliography,superscriptaddress]{revtex4-1}
\usepackage{amsmath}
\usepackage{amssymb}    %
\usepackage{amsthm}    %
\usepackage{bm}    %
\usepackage{epsfig}
\usepackage{subfigure}
\usepackage{float}
\usepackage{verbatim} 
\usepackage{pstricks}
\usepackage{color}
\usepackage{slashed}
\usepackage{multirow}
\usepackage{pstricks}
\usepackage{booktabs}
\usepackage{subfigure}
\usepackage{epstopdf}
\usepackage{braket}
\usepackage{blkarray}

\newcommand{\ssec}[1]{{\bf \textit{#1}}}
\renewcommand{\sec}[1]{\textbf{#1}\ }

\allowdisplaybreaks[3]

\def \bal#1\eal  {\begin{align} #1 \end{align}}
\newcommand{\be} {\begin{equation}}
\newcommand{\ee} {\end{equation}}

\newcommand{\ud} {\mathrm{d}}

\newcommand{\mc} {\mathcal}
\newcommand{\mb} {\mathbb}
\newcommand{\mbf} {\mathbf}

\newcommand{\ai}{{\alpha}}

\newcommand{\epi}{\epsilon}

\newcommand{\mcT} {\mathcal T}
\newcommand{\mcQ} {\mathcal Q}
\newcommand{\mcM} {\mathcal M}

\newcommand{\bfS} {{\bf S}}
\newcommand{\bfM} {{\bf M}}

\begin{document}

\hfill {\footnotesize USTC-ICTS/PCFT-21-01}

\title{Positivity in Multi-Field EFTs}
\author{Xu Li}
\email{lixu96@ihep.ac.cn}
\affiliation{
Institute for High Energy Physics, and School of Physical Sciences, University
of Chinese Academy of Sciences, Beijing 100049, China
}
\author{Hao Xu}
\email{haoxu@mail.ustc.edu.cn}
\affiliation{ Interdisciplinary Center for Theoretical Study, University of
Science and Technology of China, Hefei, Anhui 230026, China}
\affiliation{Peng Huanwu Center for Fundamental Theory, Hefei, Anhui 230026, China}
\author{Chengjie Yang}
\email{yangchengjie@ihep.ac.cn}
\affiliation{
Institute for High Energy Physics, and School of Physical Sciences, University
of Chinese Academy of Sciences, Beijing 100049, China
}
\author{Cen Zhang}
\email{cenzhang@ihep.ac.cn}
\affiliation{
Institute for High Energy Physics, and School of Physical Sciences, University
of Chinese Academy of Sciences, Beijing 100049, China
}
\affiliation{Center for High Energy Physics, Peking University, Beijing 100871, China}
\author{Shuang-Yong Zhou}
\email{zhoushy@ustc.edu.cn}
\affiliation{ Interdisciplinary Center for Theoretical Study, University of
Science and Technology of China, Hefei, Anhui 230026, China}
\affiliation{Peng Huanwu Center for Fundamental Theory, Hefei, Anhui 230026, China}

\begin{abstract} 
We discuss the general method for obtaining full positivity bounds on
multi-field effective field theories (EFTs).  While the leading order forward
positivity bounds are commonly derived from the elastic scattering of two
(superposed) external states, we show that for a generic EFT containing 3 or
more low-energy modes, this approach only gives incomplete bounds.  We then
identify the allowed parameter space as the dual to a spectrahedron,
constructed from crossing symmetries of the amplitude, and show that
finding the optimal bounds for a given number of modes 
is equivalent to a geometric problem: finding the extremal
rays of a spectrahedron. We show how this is done analytically for simple
cases, and numerically formulated as semidefinite programming (SDP) problems
for more complicated cases.  We demonstrate this approach with a number of
well-motivated examples in particle physics and cosmology, including EFTs of
scalars, vectors, fermions and gravitons.  In all these cases, we find that the
SDP approach leads to results that either improve the previous ones or are
completely new. We also find that the SDP approach is numerically much more
efficient.
\end{abstract} 

\maketitle

\sec{Introduction}
Positivity bounds are constraints on the Wilson coefficients of an effective
field theory (EFT) that can be bootstrapped from fundamental properties of the
S-matrix of the UV theory \cite{Adams:2006sv, Pham:1985cr,
Ananthanarayan:1994hf}.  Recently, there has been a lot of interest in
extending the strength and scope of the positivity bounds \cite{Zhang:2020jyn,
deRham:2017avq, deRham:2017zjm, Tolley:2020gtv, Caron-Huot:2020cmc,
Arkani-Hamed:2020blm, Bellazzini:2020cot, Vecchi:2007na, Manohar:2008tc,
Nicolis:2009qm,  Bellazzini:2015cra, Bellazzini:2016xrt, Cheung:2016wjt,
Bellazzini:2019xts, Alberte:2020jsk, Hebbar:2020ukp}, as well as applying the
bounds to constrain EFTs in various contexts (see for example
\cite{Zhang:2018shp, Bi:2019phv, Yamashita:2020gtt, Fuks:2020ujk,
	deRham:2017imi, deRham:2018qqo, Wang:2020jxr, Wang:2020xlt,
	Distler:2006if, Remmen:2019cyz, Cheung:2016yqr, Bellazzini:2017fep,
	Bonifacio:2016wcb, Bellazzini:2017bkb, Bonifacio:2018vzv,
	Bellazzini:2018paj, Melville:2019wyy, deRham:2019ctd, Alberte:2019xfh,
	Alberte:2019zhd, Chen:2019qvr, Huang:2020nqy, Tokuda:2020mlf,
	Remmen:2020vts, Trott:2020ebl, Guerrieri:2020bto, Gu:2020thj,
Gu:2020ldn, Trott:2020ebl, Bonnefoy:2020yee, Herrero-Valea:2020wxz,
Herrero-Valea:2019hde}).  In many
situations, and particularly for constraining the parameter space of the
Standard Model Effective Field Theory (SMEFT) \cite{Zhang:2018shp, Bi:2019phv,
Yamashita:2020gtt, Zhang:2020jyn, Bellazzini:2018paj, Remmen:2019cyz,
Remmen:2020vts}, the leading positivity bounds for the $s^2$ terms ($s,t$ being
the standard Mandelstam variables) in the amplitude are phenomenologically the
most relevant ones. The most widely used positivity bounds so far are based on
the forward ($t=0$) elastic scattering of two factorized states, each of which
can be an arbitrary mixture of various particle modes.  However, it has been
shown that this approach does not always give the best bounds
\cite{Zhang:2020jyn}.  In addition, obtaining the complete set of superposed
elastic bounds is known to be NP-hard \cite{Cheung:2016yqr}.

In this letter, we will establish a geometric method for obtaining the full set
of leading forward positivity bounds for EFTs with multiple low-energy
modes. It applies not only to the SMEFT, but also to all other EFTs that
involve multiple particles or multiplet particles.
We will compare with the previous results and show how the new/non-elastic
bounds arise from scattering entangled states.

\sec{Notations}
We will use capital calligraphy letters to denote rank-4 tensors
(e.g.~$\mcT\in\mb R^{n^4}$). The inner product of tensors is defined by
$\mcT_1\cdot\mcT_2\equiv\sum_{ijkl}\mcT_1^{ijkl}\mcT_2^{ijkl}$.  
We say that
$\mcT$ is positive semidefinite (PSD) if $\mcT^{ijkl}$ is a PSD matrix when
$ij$ is viewed as one index and $kl$ another, which is denoted by
$\mcT\succeq0$. The null space of this matrix is denoted as Null($\mcT$).
$\bfS_+^{n\times n}$ is the set of $n\times n$ PSD matrices.  We denote by
$\overrightarrow\bfS^{n^4}$ the set of rank-4 $n$-dimensional tensors, $\mcT$,
that satisfy the following crossing symmetries
\begin{flalign}
	\mcT^{ijkl}=\mcT^{ilkj}=\mcT^{kjil}=\mcT^{jilk}\,.
\end{flalign}
$\mcT^{i(j|k|l)}\equiv \mcT^{ijkl}+\mcT^{ilkj}$.
The set of extremal rays (ERs) of some convex cone $\mbf X$ is denoted as
ext($\mbf X$).  An ER is an element of $\mbf X$ that cannot be split into two
linearly-independent elements inside $\mbf X$. 

We shall consider the $t\to 0$ limit of a two-to-two amplitude,
$\bfM_{ij\to kl}(s)=\bfM_{ij\to kl}(s,t=0)$, which is only a function of $s$, and we
define the $\mcM$ tensor
\begin{flalign}
\label{Mdisp0}
	\mcM^{ijkl}\equiv \lim_{s\to0}\frac{d^2}{ds^2}\bfM_{ij\to kl}(s)  .
\end{flalign}
Here $i,j,k,l$ are indices for the low energy degrees of freedom, enumerating
particle species, polarization and other quantum numbers.  
We will simply call this $\mcM$ tensor ``amplitude''.

\sec{Dispersion relation} 
Axiomatic principles of the UV amplitude, such as analyticity, unitarity and
crossing symmetry, lead to a dispersion relation which expresses $\mcM^{ijkl}$
in terms of an integral of the discontinuity of the amplitude along the
positive real $s$ axis (see e.g.~\cite{Zhang:2020jyn})
\begin{equation}
\label{Mdisp}
\mcM^{ijkl}=\int_{(\epsilon\Lambda)^2}^{\infty}\! 
	\frac{\ud \mu\,{\rm Disc} \bfM_{ij\to kl}(\mu)}{2i\pi\mu^3}
	+(j\!\leftrightarrow\! l) + c.c.  
\end{equation}
where $(j\!\leftrightarrow\! l)$ denotes the previous term with $j$ and $l$
swapped. This assumes that a self-conjugate particle basis is chosen, which is
always possible by replacing $\ket{i}$ and $\ket{\bar i}$ by
$(\ket{i}+\ket{\bar i})/2$ and $(\ket{i}-\ket{\bar i})/(2i)$.  
$\epi\Lambda$ is the subtraction scale for improved positivity, below which the
EFT is valid:  we have slightly changed the definition of $\mcM^{ijkl}$
by subtracting the dispersive integral below $\epi\Lambda$,
see more explanations in Ref.~\cite{Zhang:2020jyn}.
Upon using the generalized optical theorem, this relation
implies that $\mcM^{ijkl}$ is a convex cone generated from positive linear
combinations of elements of the form
$m^{ij}m^{kl}+m^{il}m^{kj}$ \cite{Zhang:2020jyn}, i.e.,
\begin{flalign}
	\mbf C^{n^4}=\mathrm{cone}\left(\left\{
		m^{i(j}m^{|k|l)}, m\in \mb R^{n^2} \right\}\right)
\label{eq:cone}
\end{flalign}
The elements of $\mbf C^{n^4}$ are invariant under $(j\leftrightarrow l)$ and
$(i\leftrightarrow k)$ exchanges.  We will also assume that $m^{ij}$ is either
symmetric or antisymmetric, which is simply Bose symmetry for scalars, but
implies parity-conservation for vectors.  This is equivalent to further
requiring $\mbf C^{n^4}\subset \overrightarrow\bfS^{n^4}$. 

Positivity bounds arise as the boundary of $\mbf C^{n^4}$.
All components of $\mcM$ can be computed in terms of Wilson coefficients,
so bounds on $\mcM$ can be converted to bounds on these coefficients.
Conventionally, these bounds are derived by the elastic scattering of a
pair of factorized but arbitrarily superposed states, $\ket{u}=\sum_i u_i\ket{i}$
and $\ket{v}=\sum_i v_i\ket{i}$: $u^i v^j u^k v^l\mc M^{ijkl}\ge0$, thanks to
$u^iv^ju^kv^l m^{i(j}m^{|k|l)} =2(u^im^{ij}v^j)^2\ge0$.  They constrain the
signs of the elastic components in $\mcM^{ijkl}$, and also set upper and lower
bounds on inelastic scattering amplitudes
\cite{Bi:2019phv,Remmen:2019cyz,Yamashita:2020gtt,Remmen:2020vts,Trott:2020ebl}.
We will however show that these bounds are non-optimal.

The goal of this work is to understand the exact boundary of
$\mbf C^{n^4}$, which is in general beyond superposed elastic bounds.
In the presence of sufficient symmetries in the theory, an efficient way 
to do this is through the extremal positivity
approach presented in Refs.~\cite{Zhang:2020jyn}, which determines the ERs of
$\mbf C^{n^4}$ using the symmetries of the EFT, and constructs $\mbf C^{n^4}$
from the ERs (see \cite{Bellazzini:2014waa} for similar ideas). 
However, if operators that involve states not connected by any symmetries are
considered, or if the theory possesses no symmetry at all, the number of ERs
can become infinite, and this approach may not apply \cite{Yamashita:2020gtt}.
In this work, we propose a more general approach that does not rely on the
symmetries of the theory, and is thus immediately applicable to all multi-field
EFTs.

\sec{General bounds from spectrahedron} 
Let us briefly outline this general approach. First, notice that because cone $\mbf C^{n^4}$ is convex, the dual cone of $\mbf C^{n^4}$, defined as
\begin{flalign}
	{\mbf C^{n^4}}^* = \{\mcQ\ |\ \mcQ\cdot \mcM\ge0,\ \forall \mcM\in \mbf C^{n^4}\}\,,
	\nonumber
\end{flalign}
is also convex and all bounds $\mc Q\cdot \mcM\ge0$ for all $\mc Q\in {\mbf C^{n^4}}^*$
exactly describe the original cone $\mbf C^{n^4}$. 
That is, the dual of dual cone ${\mbf C^{n^4}}^*$ equals to the original cone $\mbf C^{n^4}$.
Therefore, instead of finding the $\mbf C^{n^4}$ cone of amplitudes $\mcM$, we can equivalently work with the dual cone ${\mbf C^{n^4}}^*$. To determine salient cone ${\mbf C^{n^4}}^*$, we can simply find all its ERs, as positive linear combinations of these ERs generate the whole ${\mbf C^{n^4}}^*$ \cite{KM}. 

More precisely, since $\mbf C^{n^4}$ is contained in the $\overrightarrow{\mbf
S}^{n^4}$ subspace, it is convenient to define the duality within
$\overrightarrow{\mbf S}^{n^4}$:
\begin{flalign}
	{\mbf Q^{n^4}}\equiv{\mbf C^{n^4}}^* = \{\mcQ\in\overrightarrow{\mbf S}^{n^4}\ |\ \mcQ\cdot \mcM\ge0,\ \forall \mcM\in \mbf C^{n^4}\}\,,
	\nonumber
\end{flalign} 
We now need to find
$\mbf Q^{n^4}$. For any $\mc Q\in \mbf Q^{n^4}$,
$\mc Q\cdot \mcM\ge0\Leftrightarrow \mcQ^{ijkl}m^{i(j}m^{|k|l)}\ge0
\Leftrightarrow 2\mcQ^{ijkl}m^{ij}m^{kl}\ge0$ for any $m$ (thanks to 
$\mc Q\in \overrightarrow\bfS^{n^4}$), which is equivalent to $\mcQ\succeq0$. 
Therefore we have
$
\mbf Q^{n^4}=\mbf
S_{+}^{n^2\times n^2} \cap \overrightarrow\bfS^{n^4}
$
which is known
as a spectrahedron.  Geometrically, a spectrahedron is the intersection
of the cone of PSD matrices ($\mbf S_+^{n^2\times n^2}$) with an
affine-linear space (in our case, $\overrightarrow\bfS^{n^4}$), and is a well
studied geometric object, intimately linked to SDP---the latter is simply an
optimization on a spectrahedron \footnote{for applications of SDP in conformal
	bootstrapping, see \cite{Poland:2018epd} and references therein.}. The complete and independent positivity bounds are simply $\mcQ\cdot\mcM\ge0$
for all $\mcQ\in\mathrm{ext}( \mbf Q^{n^4} )$.

We have essentially converted the problem of finding positivity bounds to a
geometric problem: finding the ERs of a {\it spectrahedron}.  Note that these
ERs are in the dual space $\mbf Q^{n^4}$, and are to be distinguished from the
ERs of the physical amplitude space $\mbf C^{n^4}$. The latter have been used
in Refs.~\cite{Zhang:2020jyn} to directly construct the boundary of $\mbf
C^{n^4}$.  As we have mentioned, this procedure becomes cumbersome to use for
theories with large $n$ but insufficient symmetries to determine the ERs. On
the contrary, we will see that the new approach presented here does not have
this limitation.

How do we search for the ERs in $\mbf Q^{n^4}$? Just like a polyhedron, a
spectrahedron has many (flat) faces of different dimensions. It has been shown in
Ref.~\cite{ramana} that for any point $\mcQ$ in a spectrahedron, there exists a
unique face $F(\mcQ)$ that contains $\mcQ$ with the lowest
possible dimension and where Null$(\mcQ)$ is constant (independent of
where $\mcQ$ is on face $F(\mcQ)$).  This provides a characterization of the faces,
and in particular the ERs (which are 1-dimensional faces) of a spectrahedron.
Let $u_1,u_2,\cdots u_k$ be a basis of Null$(\mcQ)$ and $\mcQ_1,\mcQ_2,\dots,
\mcQ_m$ be a basis of $\overrightarrow\bfS^{n^4}$, then the null space of the
following $(n^2k)\times m$ matrix
\begin{flalign}
	B=\begin{bmatrix}	
\mcQ_1 u_1 & \cdots & \mcQ_m u_1\\
\vdots & \ddots & \vdots \\
\mcQ_1 u_k & \cdots & \mcQ_m u_k
	\end{bmatrix}\,,
	\label{eq:bmatrix}
\end{flalign}
gives the linear subspace that contains $F(\mcQ)$.  If Null($B$) is 1-dimensional, then
$\mcQ$ is an ER.  
The positivity bounds are simply $\mc Q\cdot \mc M\ge0$ for all such $\mc Q$'s.

\sec{Toy model:~multi-scalar}
Consider an EFT of $n$ scalar modes $\phi_{i=1,...,n}$.  At the tree-level, the relevant operators are dim-8,
and a basis can be chosen as
	$O_{ijkl}=\partial_\mu\phi_i \partial^\mu\phi_j \partial_\nu\phi_k
\partial^\nu\phi_l$, which has symmetry $O_{ijkl}=O_{jikl}=O_{ijlk}=O_{klij}$.
Let us consider simply a $Z_2$ symmetric model ($\phi_i \to -\phi_i$). 
The amplitude can be computed straightforwardly. We find
$\mcM_{iiii}=4C_{iiii}$,
$\mcM_{iijj}=\mcM_{ijji}=\mcM_{jiij}=\mcM_{jjii}=C'_{iijj}\equiv
C_{iijj}+\frac{1}{2}C_{ijij}$, and $\mcM_{ijij}=\mcM_{jiji}=C_{ijij}$. All
other elements vanish. 

The same $Z_2$ symmetry can be applied to its dual space, the spectrahedron
$\mbf Q^{n^4}$. For $n=2$,
a general element in $\mbf Q^{n^4}$ can be parameterized as:
\begin{flalign}
\mc Q=\begin{bmatrix} 
x_1 & x_2 & & \\
x_2 & x_3 & & \\
& & x_4 & x_2   \\
& & x_2 & x_4 \\
\end{bmatrix}, x_{1,3}\ge0,\ x_1x_3\ge x_2^2,\ x_4\ge |x_2|
\nonumber
\end{flalign}
where the rows (columns) correspond to the
$i,j$ ($k,l$) pairs taking $(1,1),(2,2),(1,2),(2,1)$. 
The $2\times2$ block-diagonal structure is due to the $Z_2$ symmetry.
Crossing symmetry is reflected the common matrix elements,
while $\mc Q\succeq0$ leads to the inequalities.
Writing $\mc Q\equiv x_i\mc Q_i$, 
each $\mc Q$ can be represented by a $\vec x=(x_1,\dots,x_4)$.
From these inequalities, we can find the ERs:
$\vec x_{e1}(r)=(1,r,r^2,|r|)$ and $\vec x_{e2}=(0,0,0,1)$, where
$r$ is an arbitrary real number, and $\vec x_{e1}(r)$ is extremal
for any $r$.
They are complete because any other $\vec x$ can be written as
$\vec x=\frac{x_2^2}{x_3}\vec x_{e1}(\frac{x_3}{x_2})
+(x_1-\frac{x_2^2}{x_3})\vec x_{e1}(0)+(x_4-|x_2|)\vec x_{e2}$,
which is a positively weighted sum.

Each ER corresponds to an independent positivity bound.
The second ER, $x_{e2,i}\mcQ_i\cdot \mcM\ge0$, simply gives
$C'_{1212}\ge0$.
The $r$-dependent ER, $\vec x_{e1}(r)$, gives
$4C_{2222}r^2+4C'_{1122}r+2C_{1212}|r|+4C_{1111}\ge0$.
Together they are equivalent to:
\begin{flalign}
	&C_{1111}\ge0,\ C_{2222}\ge0,\ C_{1212}\ge0\\
	&4\sqrt{C_{1111}C_{2222}}\ge \pm(2C_{1122}+C_{1212})-C_{1212}
	\label{eq:bounds1}
\end{flalign}
As a quick application of this result, it improves the previous positivity
bounds on the parameters of the Higgs-Dilaton inflationary model
\cite{Herrero-Valea:2019hde}.

\begin{figure}[htb]
	\begin{center}
		\includegraphics[width=.9\linewidth]{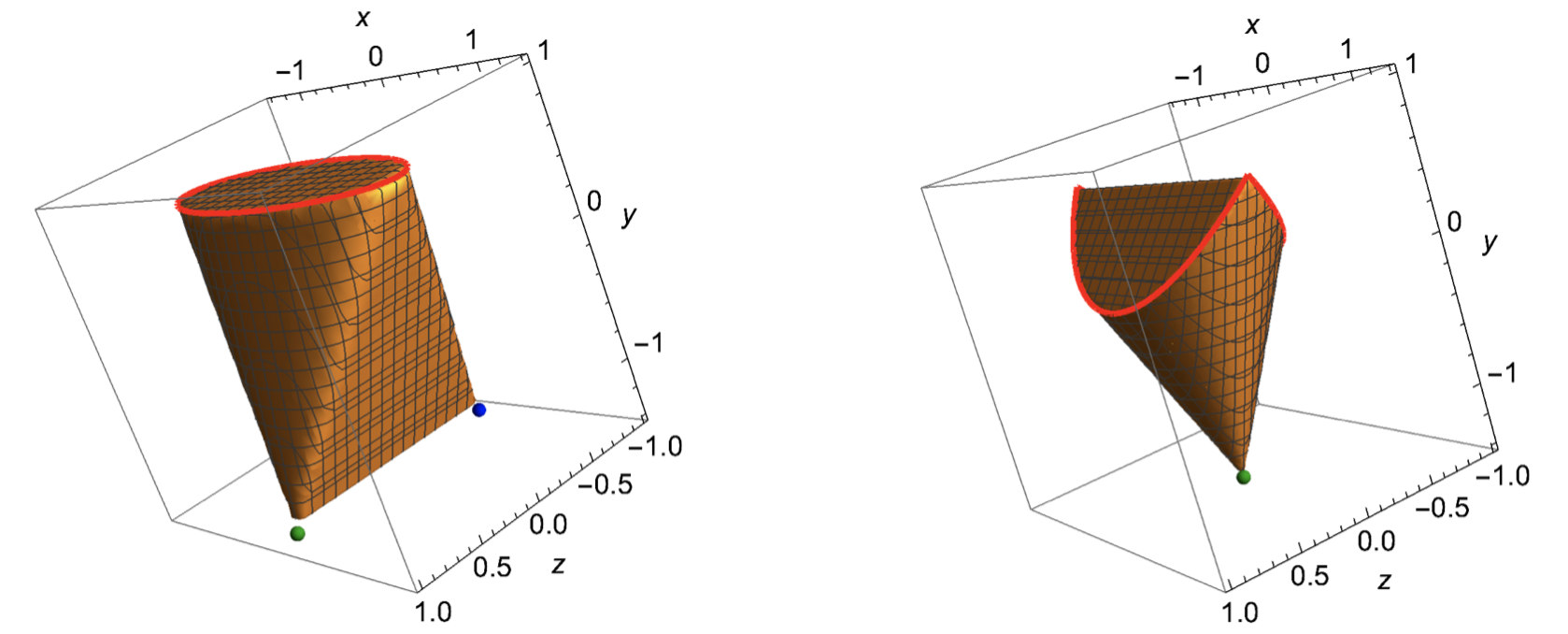}
	\end{center}
	\caption{
	3-dimensional slice of $\mbf C^{2^4}$ (left) and $\mbf
	Q^{2^4}$(right) for the bi-scalar toy example with $Z_2$ symmetry.
	The three axes in the left plot are taken to be {\footnotesize $(x,y,z)=
	\left( 
	2\sqrt{6}\left( C_{1111}-C_{2222} \right),
	\sqrt{2}\left( 2C_{1111}-C_{1212}+2C_{2222} \right),
	\sqrt{3} C'_{1122} 
	\right)$},
	normalized to $4C_{1111}+C_{1212}+4C_{2222}=1$.
	Those in the right plot are the same but with $C^{(}{}'{}^{)}_{ijkl}\to
	Q_{ijkl}$.
	}
	\label{fig:QM}
\end{figure}

To illustrate the relation between $\mbf C^{2^4}$ and its dual, in
Figure~\ref{fig:QM} we display the 3D cross sections of the physical amplitudes
$\mbf C^{2^4}$ and the spectrahedron $\mbf Q^{2^4}$, which are both 4D cones.
The two types of ERs of $\mbf Q^{2^4}$ are highlighted by the red and green
extreme points, respectively. The boundary of the $\mbf C^{2^4}$ are dual to
these ERs: a vertex in $\mbf Q^{2^4}$ corresponds to a facet in $\mbf C^{2^4}$
and vice versa, as implied by duality. Finding the full bounds is
therefore equivalent to finding ext($\mbf Q^{2^4}$).
On the other hand, the ERs of the physical amplitudes $\mbf C^{2^4}$
are also highlighted. They can be of special physical interest,
and we refer to Ref.~\cite{Fuks:2020ujk} for potentially interesting
phenomenological consequences. (More general cases with more modes and without $Z_2$ symmetry are presented in Supplementary Material.)

Our approach always gives the complete bounds
available from the dispersion relation. In contrast, the conventional
positivity approach based on elastic scattering can be incomplete for a model with multiple modes. The elastic bounds are complete iff all elements of $\mathrm{ext}(\mbf Q^{n^4})$ can be written in  form of $\mcQ_{uv}^{ijkl}\equiv u^iv^ju^kv^l+v^iu^jv^ku^l$. This can always be done for bi-scalar models, even without the $Z_2$ symmetry (see Supplementary Material). However,  this ceases to be true when there are 3 and more scalars. 
To see this, it suffices to give an example of $\mcQ$ being extremal in $\mbf
Q^{3^4}$ but not of the form of $\mcQ_{uv}$. One explicit example is $\mcQ_{\rm ex}=\sum_{\alpha=1}^4
U_\alpha^{ij}U_\alpha^{kl}$, with the following four $U_\alpha$ matrices:
\begin{flalign}
	\begin{bmatrix}
		1 &0&0\\
		0&0&0\\
		0&0&1
	\end{bmatrix},
	\begin{bmatrix}
		0 &0&1\\
		0&1&0\\
		1&0&1
	\end{bmatrix},
	\begin{bmatrix}
		0 &1&1\\
		1&0&1\\
		1&1&1
	\end{bmatrix},
	\begin{bmatrix}
		0 &1&1\\
		-1&0&0\\
		-1&0&0
	\end{bmatrix}
	\label{eq:U}
\end{flalign}
$\mc Q_{\rm  ex}$ is a rank-4 matrix, so it cannot be written as some $\mcQ_{uv}$,
which is at most rank-2 by definition.
We will explain the physics interpretation of $\mc Q_{\rm ex}$ later,
using the SM flavor operators as an example.

We see that in the most general case, {\it elastic positivity is incomplete for
EFTs with more than 2 low-energy modes.} 
In practice, however, the existence of symmetry relations can delay the
appearance of non-elastic bounds. 
For example, the 4-$W$ operators presented in Ref.~\cite{Zhang:2020jyn} contain non-elastic bounds. The $W$-boson carries 2 helicities and is charged under the adjoint of SU(2), which is equivalent to the fundamental of SO(3), thus the number of independent components in this case is 6. However, if reducing the SO(3) to SO(2), which leads to 4 independent components left, there is no non-elastic bound any more. 

\sec{General numerical method}
For a model with many low energy modes, the optimal positivity bounds can be efficiently obtained numerically. To see this, note that $\mc M$ being in $\mbf C^{n^4}$ is equivalent to $\mcQ\cdot \mcM\ge 0$ for all $\mcQ\in
\mbf Q^{n^4}$. This means we can get the optimal bounds by requiring the following semi-definite program (SDP)
\begin{flalign}
&\mbox{min}\quad \mcQ\cdot \mcM \nonumber\\
&\mbox{subject to}\quad \mcQ\in \mbf Q^{n^4}
\end{flalign}
has a non-negative minimum.  This solves the problem in polynomial time complexity,
and always gives the best bounds within given numerical accuracy, in contrast to
the elastic positivity approach, which is NP-hard and leads to incomplete bounds.

It is sometimes useful to explicitly describe the boundary of $\mbf C^{n^4}$.
To this end, an MC approach can be adopted in order to obtain a random sampling
of linear bounds.  To find an ER, one simply:
\begin{enumerate}
	\item Pick a random point $\mcQ$ in $\mbf Q^{n^4}$,
		and compute $F(\mcQ)$ using Eq.~(\ref{eq:bmatrix}).
	\item If $F(\mcQ)$ is 1-dimensional, then $\mcQ$ is on an ER; 
		otherwise, take a random straight line in $F(\mcQ)$, and find its
		intersection(s) with the boundary of $\mbf Q^{n^4}$ (which is an
		SDP problem).
	\item Let $\mcQ$ be one of the intersection points and iterate, until
		an ER is found.
\end{enumerate}
The iteration will take $\mcQ$ to a random ER.  If the problem only has a
finite number of bounds,
this iteration will capture all bounds. This is often the case if one
considers the self-interactions of some multiplet particle (see examples in
Ref.~\cite{Zhang:2020jyn,Trott:2020ebl}). For non-polyhedral cones, we will
get a sampling of bounds with a finite number of iterations.

Our new approach in principle captures all the information from the
forward and twice-subtracted dispersion relation, and improves many previous
results based on elastic scattering.  {\it We now demonstrate this in subspaces of SMEFT.}

\sec{SM gauge bosons}
In the SMEFT, positivity bounds at dim-8 on gauge-boson operators are partially known
\cite{Zhang:2018shp, Bi:2019phv,
Remmen:2019cyz,Zhang:2020jyn,Yamashita:2020gtt,Trott:2020ebl}.  To test our new
approach, we consider parity-conserving 4-gluon SMEFT operators.  There are 6
relevant dim-8 operators (defined in Ref.~\cite{Murphy:2020rsh}; see also Supplementary Material), schematically of the form $\mbf G^4$. The dim-6
operator $O_{G}=f^{ABC}G^{A\nu}_{\mu}G^{A\rho}_{\nu}G^{A\mu}_{\rho}$ can also
contribute through diagrams with two insertions.  The amplitude $\mcM$
can then be mapped to
\begin{flalign}
\vec c\equiv\begin{bmatrix}
	C_{G^4}^{(1)} & C_{G^4}^{(2)} & C_{G^4}^{(3)} & C_{G^4}^{(4)} & C_{G^4}^{(7)} & C_{G^4}^{(8)} & c_G^2
	\label{eq:gluonCs}
\end{bmatrix}
\end{flalign}
where $C_{G^4}^{(i)}$ is the coefficient of $Q_{G^4}^{(i)}$ defined in
Ref.~\cite{Murphy:2020rsh}, and $c_G$ is the coefficient of $O_G$.

Using the MC approach, we find 45 linear inequalities, which we have also
verified with the symmetric extremal approach
\cite{Zhang:2020jyn}. They can be written in the form of $\vec x\cdot
\vec c\ge0$, and the first 6 $\vec x$ vectors are
\begin{flalign}
\begin{aligned}
&[0,0,0,1,0,0,0]\\ 
&[0,2,0,1,0,0,0]\\ 
\end{aligned}\quad
\begin{aligned}
&[0,0,1,1,1,0,0]\\ 
&[0,0,3,0,2,0,0]\\ 
\end{aligned}\quad
\begin{aligned}
&[2,0,1,0,0,0,0]\\ 
&[0,0,0,3,0,2,0]\\ 
\end{aligned}
\nonumber
\end{flalign}
while the rest 39 are given in Supplementary Material.  Previous results
on parity-conserving operators based on selected elastic scattering in 
Ref.~\cite{Remmen:2019cyz} can be reproduced already by the 3rd to the 6th
$\vec x$ vectors.
We emphasize that this is a new result and an important step
towards the full set of SMEFT positivity bounds.

The new approach is most powerful when multiple gauge-boson fields are
incorporated, where the positivity cone is no longer polyhedral. A
phenomenologically relevant case is the operators that characterize the
anomalous quartic-gauge-boson couplings (QGCs), which is an essential part of
the electroweak program at the LHC (see
Refs.~\cite{Sirunyan:2019der,CMS:2020meo,Sirunyan:2020tlu} for recent
results).  Knowing positivity
bounds for these operators will provide guidance for future experimental
searches.  For operators sourcing only the transversal modes,
using the SDP approach, we find that the coefficient space is cut down to
$0.681\%$ of the total.  This agrees with Ref.~\cite{Yamashita:2020gtt}, where
the same number is obtained by approximating the amplitude space by a
polyhedral cone with a large number ($N\approx \mathcal{O}(10^3))$ of edges
and extrapolating $N\to \infty$, which is much less efficient.  The full set of
aQGC bounds can also be determined by the SDP approach. We will present it in a future
work.

\sec{SM flavor sector}
A perhaps more relevant example is the SMEFT operators in the
flavor sector.  The SM fermions come with 3 generations, so full positivity
bounds cannot be derived from elastic scattering of mixed flavors; flavor
symmetry needs not be a symmetry of the SMEFT, so the symmetric extremal approach
\cite{Zhang:2020jyn} does not apply. The SDP approach solves this problem.
Consider one fermion species $f$, say the right-handed electron $f=e_R$, but
for all 3 generations. Using the Fierz identity, the dim-8 four-fermion operators
can always be written as $O_{ijkl}=\partial_\mu(\bar f_i\gamma_\nu f_j)
\partial^\mu(\bar f_k\gamma^\nu f_l)$, where $i,j,k,l$ are flavor indices.
Since $O_{ijkl}=O_{klij}$, we only count the independent ones. 
The crossing symmetric amplitude $\mcM$ depends on 21 independent
Wilson coefficients (see Supplementary Material).

\begin{figure}[htb]
	\begin{center}
		\includegraphics[width=\linewidth]{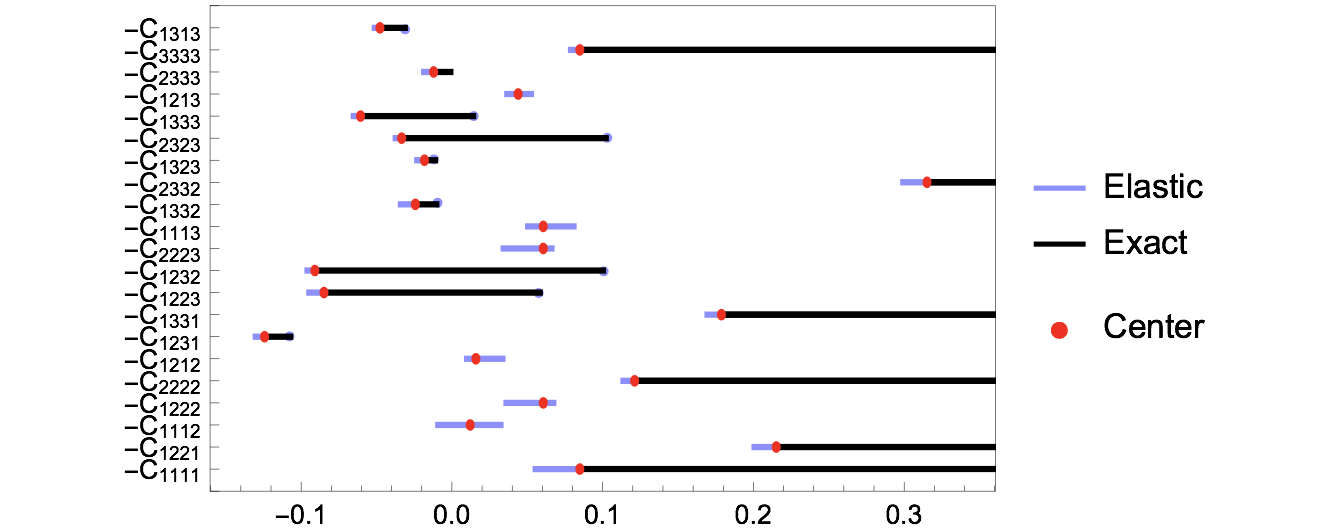}
	\end{center}
	\caption{Comparison of the elastic bounds (``Elastic'') and the SDP
		bounds (``Exact''). The red dots (``Center'') denote the
		set of coefficients $\vec C_0$ which saturates the non-elastic
		bound $\mcQ_{\rm ex}$ given in Eq.~(\ref{eq:U}).
	}
	\label{fig:flavor}
\end{figure}

To illustrate the improvement of the new approach, we pick a set of
coefficients $\vec C_0$ that saturates the non-elastic bound $\mcQ_{\rm ex}$ given in
Eq.~(\ref{eq:U}), and display both elastic and the exact bounds in
Figure~\ref{fig:flavor}.  These bounds are obtained by varying one operator
at a time, while keeping the others fixed at $\vec C_0$, whose
values are indicated with red dots. 
Elastic amplitudes are only
bounded from below, while others are bounded from both sides.
Since $\vec C_0$ is chosen to saturate the
$\mcQ_{\rm ex}$ bound, the exact bounds could often uniquely fix the coefficients, so some exact bounds are not visible in the plot.

The new bound from $Q_{\rm ex}$ can be interpreted as coming from 
combining four channels between initial and final states
$\ket{I_\alpha}=\ket{F_\alpha}=U_\alpha^{ij}\ket{i}\otimes\ket{j}$,
for $\alpha=1,2,3,4$. The $U$ matrices are given in Eq.~(\ref{eq:U})
and are at least rank-2, implying the two incoming particles are entangled.
The $U_1$ matrix, for example, describes the scattering
of the entangled state $\ket{I_1}=\ket{F_1}=\ket{e}\otimes \ket{e}
+\ket{\tau}\otimes \ket{\tau}$.
Individually, these states cannot be used to construct positivity bounds,
because the $u$-channel contribution in the dispersion relation,
$U_\ai^{ij}U_\ai^{kl}m^{il}m^{kj}$, is not positive semidefinite. However,
the $\mcQ_{\rm ex}$ tensor combines these channels together such that $\sum_{\alpha=1}^4
U_\alpha^{ij}U_\alpha^{kl} \in \overrightarrow{\mbf S}^{n^4}$ is
crossing-symmetric, which guarantees that both $s$- and $u$-channels are
positive.

Positivity bounds for the flavor operators of the SMEFT are phenomenologically
relevant, 
as the existence of flavor-violating effects (e.g.~$\mu\to 3e$) would
set lower bounds on the flavor-conserving ones (e.g.~$e^+e^-\to e^+e^-$),
providing important guidance for future experiments \cite{Remmen:2020vts}.
While dim-6 contributions potentially give the dominant contribution, future
precision measurements are likely to have sufficient precision to simultaneously
determine both dim-6 and dim-8 effects through global fits \cite{Fuks:2020ujk}.
Novel observables have also been designed to extract dim-8 information without
being affected by the dim-6 ones \cite{Alioli:2020kez}. Phenomenological studies for dim-8
SMEFT have started in the recent years
\cite{Ellis:2017edi,Ellis:2018cos,Bellazzini:2018paj,Ellis:2019zex,Ellis:2020ljj,
Gu:2020ldn,Corbett:2021eux,Hays:2018zze,Bellazzini:2017bkb,Fuks:2020ujk,Alioli:2020kez},
and their interplay with
positivity bounds may reveal crucial information about UV physics.  Our new
approach guarantees the best positivity bounds at dim-8, and is thus crucial
for fully capturing this information.

\sec{Summary}
We have shown that the full $s^2$ positivity bounds for EFTs with $n$ low-energy
modes are given by the ERs of the spectrahedron $\mbf Q^{n^4}$. We have formulated the problem of finding the optimal bounds as a semi-definite program, which can be efficiently solved in polynomial times.
 We have presented
realistic examples and improved previous results in the areas of cosmology,
 LHC and flavor physics (see Supplementary Material for more details, with Refs.~\cite{Alberte:2019lnd, deRham:2010kj, Hilbert1888, Keltner:2015xda} included there), which are all useful physical results by
themselves.  Our approach is straightforwardly applicable to all multi-field
EFTs, and represents a crucial step towards fully extracting the positivity
constraints for realistic EFTs with many degrees of freedom. \\

\begin{acknowledgments}
{\bf Acknowledgments} We would like to thank Anna Tokareva and Zi-Yue Wang for helpful discussions and
comments.  CZ is supported by IHEP under Contract No.~Y7515540U1, and by
National Natural Science Foundation of China (NSFC) under grant No.~12035008.
SYZ acknowledges support from the starting grants from University of Science and Technology of China under grant No.~KY2030000089 and GG2030040375, and is also supported by National Natural Science Foundation of China under grant No.~11947301, 12075233 and 12047502, and supported by the Fundamental Research Funds for the Central Universities under grant No.~WK2030000036. \\
\end{acknowledgments}

\bibliography{refs}
\bibliographystyle{JHEP}

\begin{widetext}

\section{Supplemental Material}

\subsection{More details about multi-scalar EFTs}
\label{app:scalar}

Here we provide more details about the multi-scalar model we have used to illustrate the basic ideas of our new approach. Apart from illustration purposes, the multi-scalar model is a theory with little symmetry, and from the viewpoint of the positivity cones, other theories can be obtained by appropriate symmetric projections of this cone. Scalar fields are also widely used in model building in cosmology, and as briefly mentioned in the main text, these results can readily be used to improve the physical bounds in the literature.\\

\ssec{Bounds for multi-scalar EFT with $Z_2$}.  
In the main text, we have constructed the $\mbf Q^{n^4}$ cone for $n=2$ with $Z_2$ symmetry, which can be easily done by directly implementing crossing symmetry and $Z_2$ symmetry. Here we will construct the $\mbf Q^{n^4}$ cone for higher $n$. We shall explicitly illustrate the procedure for the case of $n=3$, and an explicit construction for higher $n$ is also possible but more involved.

For scalar EFTs that are invariant under $\phi_i \to -\phi_i$ for all $i$, an
element $\mcQ_{ijkl}$ of the $\mbf Q^{n^4}$ spectrahedron, which satisfies the same symmetries as $\mcM_{ijkl}$,
is a $n^2\times n^2$ block diagonal matrix, with the first block being
\begin{flalign}
b_0=\begin{bmatrix} 
\mcQ_{1111} & \mcQ_{1122} & \dots &\mcQ_{11nn} \\
\vdots & \multicolumn{2}{c}{ \ddots} & \vdots \\
    \mcQ_{11nn} & \mcQ_{22nn}&\dots  & \mcQ_{nnnn} 
    \end{bmatrix}  ,
\end{flalign}
followed by
\begin{flalign}
b_{ij}=\begin{bmatrix} 
\mcQ_{ijij} & \mcQ_{ijji} \\
\mcQ_{jiij} & \mcQ_{jiji} 
    \end{bmatrix}
    =
\begin{bmatrix} 
\mcQ_{ijij} & \mcQ_{iijj} \\
\mcQ_{iijj} & \mcQ_{ijij} 
    \end{bmatrix}
  ,\  1\le i<j\le n   .
\end{flalign}
So all off-diagonal elements in the first block, $(b_0)_{ij}$, are equal to the
off-diagonal element of $b_{ij}$. $\mcQ$ being positive semi-definite requires that all the blocks
be individually positive semi-definite.  

Using the fact that the ERs of $\bfS_+^{n\times n}$ are the rank-1
symmetric matrices, one can show that $\mbf Q^{n^4}$ has two sets of ERs,
defined by the following conditions, respectively,
\begin{flalign}
	&Q_\mathrm{ex1}(x): b_0=xx^T,\ \mbox{and}\ \mcQ_{ijij}=|\mcQ_{iijj}| ~
	\mbox{in all }b_{ij}    ,
	\nonumber\\
	&Q_\mathrm{ex2}^{ij}:
	b_0=\mathbf{0}_{n\times n}, \ 
	b_{kl}=\left\{\begin{array}{ll}
		\mathbf{1}_{2\times2} & (k,l)=(i,j)
		\\
		\mathbf{0}_{2\times2} & \mbox{otherwise} ,
	\end{array}
		\right.
	\label{eq:erset1}
\end{flalign}
where $Q_\mathrm{ex1}$ is a function of an arbitrary $n$-dimensional vector $x$, and
$Q_\mathrm{ex2}^{ij}$ depends on two integers, $1\le i<j\le n$.  One can easily
check that these are indeed extremal using Eq.~(5). (For $n=2$, we have shown that they are the only ERs.) In the following, we will prove that for $n=3$ there are no more ERs in
addition to $Q_\mathrm{ex1}(x)$ and $Q_\mathrm{ex2}^{ij}$.

First, let us set up the notation.
An element of $\mbf Q^{3^4}$ can be parameterized as
\begin{flalign}
\mcQ(\vec x)=\begin{bmatrix} 
(b_0)_{3\times3} & & & \\
 & (b_{12})_{2\times2} & & \\
& & (b_{13})_{2\times2} &    \\
& &  & (b_{23})_{2\times2} \\
\end{bmatrix}
=
\begin{blockarray}{lcccccccccr}
   kl=&11 & 22 & 33 & 12 & 21 & 13 & 31 & 23 & 32 &\\
\begin{block}{l[ccccccccc]r}
&x_{1}&y_{12}&y_{13}&&&&&&&ij=11\\
&y_{12}&x_{2}&y_{23}&&&&&&&22\\
&y_{13}&y_{23}&x_{3}&&&&&&&33\\
&&&&z_{12}&y_{12}&&&&&12\\
&&&&y_{12}&z_{12}&&&&&21\\
&&&&&&z_{13}&y_{13}&&&13\\
&&&&&&y_{13}&z_{13}&&&31\\
&&&&&&&&z_{23}&y_{23}&23\\
&&&&&&&&y_{23}&z_{23}&32\\
\end{block}
\end{blockarray}
\end{flalign}
where $\vec x =\begin{bmatrix}
	x_1&x_2&x_3&y_{12}&y_{13}&y_{23}&z_{12}&z_{13}&z_{23} \end{bmatrix}$.
A basis for $\mbf Q^{3^4}$, denoted by $\{\mc Q_i\}$, can be chosen such that
$\mcQ(\vec x)=\sum_i x_i\mc Q_i$.  For convenience we further separate
the basis into $\{\mc Q_{x,k}\}$, $\{\mc Q_{y,kl}\}$ and $\{\mc Q_{z,kl}\}$, by
writing
\begin{flalign}
	\mcQ(\vec x)=\sum_{k=1}^3 x_i\mc Q_{x,k} + \sum_{1\le k<l\le3}
	y_{kl} \mc Q_{y,kl}+ \sum_{1\le k<l\le3}
	z_{kl} \mc Q_{z,kl}  .
\end{flalign}
The null space of $\mcQ(\vec x)$ is a direct sum of the null spaces
of the diagonal blocks: $\mathrm{Null}[\mcQ(\vec x)]=\mathrm{Null}(b_0)\oplus
\mathrm{Null}(b_{12})\oplus\mathrm{Null}(b_{13})\oplus\mathrm{Null}(b_{23})$.
For a given $\vec x$,
we can construct a basis of $\mathrm{Null}\left[\mcQ\left(\vec x\right)\right]$,
denoted as $U(\vec x)=\{u_i\}$, by combining the basis vectors of each subspace.
For example, if all 4 subspaces are one-dimensional, we take
\begin{flalign}
&u_1=v^{(0)}\oplus \mbf{0}_2 \oplus \mbf{0}_2\oplus \mbf{0}_2,\qquad
u_2=\mbf{0}_3\oplus v^{(12)} \oplus \mbf{0}_2\oplus \mbf{0}_2,
\\
&u_3=\mbf{0}_3\oplus \mbf{0}_2 \oplus v^{(13)} \oplus \mbf{0}_2,\qquad
u_4=\mbf{0}_3\oplus \mbf{0}_2 \oplus \mbf{0}_2\oplus v^{(23)},
\\
&U(\vec x)=\left\{u_1,u_2,u_3,u_4\right\},
\end{flalign}
and so on, where $v^{(0)}$ and $v^{(kl)}$ are the basis vectors of
$\mathrm{Null}(b_0)$ and $\mathrm{Null}(b_{kl})$ respectively. $\mbf{0}_m$ is an
$m$-dimensional zero vector.

The necessary and sufficient condition for $\mcQ(\vec x_0)$ to be extremal
in $\mbf Q^{3^4}$ is that the $B$ matrix defined in Eq.~(5)
has a one-dimensional null space. This means the following system,
\begin{flalign}
	\mcQ(\vec x)u_j = \sum_{i} x_i\mcQ_iu_j=0 ~ ~ \mbox{for all }u_j\in U(\vec
	x_0)\,,
	\label{eq:cri}
\end{flalign}
uniquely determines $\vec x$ up to normalization. 
This requirement is simply that $\mcQ(\vec x)$ has the same null space as
$\mcQ(\vec x_0)$. In other words, if $\mcQ(\vec x_0)$ is extremal, 
$\mathrm{Null}[\mcQ(\vec x)]=\mathrm{Null}[\mcQ(\vec x_0)]$ implies $\vec
x=\lambda \vec x_0$. 

Now assume an element $\mcQ(\vec x_0)$ is an ER. Depending on the ranks
of the diagonal blocks $b_0$ and $b_{kl}$, there are several cases:
\begin{enumerate}
	\item $\mathrm{rank}(b_0)=0$, i.e.,~$b_0$ vanishes. In this case the
		only non-vanishing components in $\vec x_0$ are the
		$z_{kl}$'s, and $\mcQ(\vec x_0)\succeq0$ requires $z_{kl}\ge0$.
		Obviously, the only ERs of this type are the 
		second type in Eq.~(\ref{eq:erset1}) (i.e.,~only one $b_{kl}$
		matrix is nonvanishing. It is easy to show that $\mcQ$ of this
		form cannot be split into two other elements, as the vanishing
		$b$ blocks cannot be a sum of two different PSD matrices, and thus
		fulfilling the definition of an ER.)

		In the following we consider $\mathrm{rank}(b_0)\ge1$.
	\item $\mathrm{rank}(b_{kl})=2$ for some $1\le k<l\le 3$. The
		null space for $b_{kl}$ does not exist, and $v^{(kl)}$ vanishes.
		Consider Eq.~(\ref{eq:cri}). Since there is no $u_i$ vector
		constructed from $v^{(kl)}$, $Q_{z,kl}u_i=0$ for all $u_i$,
		and so there is no equation from (\ref{eq:cri}) that can be
		used to constrain $z_{kl}$. The only possible ER in this case
		is that all components in $\vec x_0$ vanish except for
		$z_{kl}$.  This again gives the second type in
		(\ref{eq:erset1}).

		In the following we consider $\mathrm{rank}(b_{kl})\le1$
		for all $1\le k<l\le 3$.
	\item $\mathrm{rank}(b_{kl})=0$ for all $1\le k<l\le 3$
		(i.e.,~all $b_{kl}$'s vanish).
		Each $b_{kl}$ has a two-dimensional null space,
		and so Eq.~(\ref{eq:cri}) would simply force $y_{kl}=z_{kl}=0$,
		so that Null$(b_{kl})$ will not change.
		We are left with three parameters, $x_1,x_2,x_3\ge0$.
		For $\mcQ(\vec x_0)$ to be extremal, only one of them can
		be nonzero. So we have $\mathrm{rank}(b_0)=1$, a special
		case for the first type in Eq.~(\ref{eq:erset1}).
	\item $\mathrm{rank}(b_{kl})=1$ for all $1\le k<l\le 3$,
		i.e.,~$z_{kl}=|y_{kl}|$.  Each $b_{kl}$ has a one-dimensional
		null space, and so in Eq.~(\ref{eq:cri}), the corresponding
		$v^{(kl)}$ vector, being $(1,1)^T$ or $(1,-1)^T$, will enforce
		$z_{kl}=-y_{kl}$ or $z_{kl}=y_{kl}$ through equation
		$\left(y_{kl}Q_{y,kl}+z_{kl}Q_{z,kl}\right)v^{(kl)}
		=(y_{kl}\pm z_{kl})(\pm1,1)^T=0$. We are left with 6 free
		parameters in $\vec x_0$: three $x_k$'s and three $y_{kl}$'s. To constrain
		them up to normalization, we need 5 additional independent
		equations coming from Eq.~(\ref{eq:cri}).  Since each null
		vector $v^{(0)}$ gives at most 3 independent ones, we need at
		least two null vectors, which means $\mathrm{rank}(b_{0})=1$.
		This leads to the first type in Eq.~(\ref{eq:erset1}).
%
	\item One $b_{kl}$ is rank-0 (i.e.,~it vanishes) while the other two are
		rank-1. Suppose $b_{12}$ vanishes. Similar to case 3 and case 4,
		the null vectors of all three $b_{kl}$'s give the following constraints
		through Eq.~(\ref{eq:cri}): $y_{12}=z_{12}=0$, $y_{13}=\pm
		z_{13}$, $y_{23}=\pm z_{23}$. 5 free parameters are left:
		$x_1,x_2,x_3,y_{13},y_{23}$. Again, two null vectors
		$v_{1,2}^{(0)}$ are needed, which means $\mathrm{rank}(b_{0})=1$.
	\item Two $b_{kl}$'s are rank-0 (i.e.,~they vanish) while the other one is
		rank-1. Suppose $b_{13}$ and $b_{23}$ vanish. Similar to case 3 and
		case 4, we have the following constraints: $y_{12}=\pm z_{12}$,
		$y_{13}=z_{13}=y_{23}= z_{23}=0$. 4 free parameters are left:
		$x_1,x_2,x_3,y_{12}$. The $b_0$ block is divided into two
		smaller blocks,
		$b_{01}=\begin{bmatrix}x_1&y_{12}\\y_{12}&x_2\end{bmatrix}$ and
		$b_{02}=\begin{bmatrix}x_3\end{bmatrix}$.  If both are
		nonvanishing, $\mcQ(\vec x_0)$ cannot be extremal, because
		$x_3\to 2x_3$ does not change Null$(b_0)$, which means
		Eq.~(\ref{eq:cri}) cannot constrain $\vec x_0$ up to
		normalization; 
		If $b_{01}$ vanishes, we are back to case 3;  If
		$b_{02}$ vanishes, we are left with 3 free parameters,
		$x_1,x_2,y_{12}$. At least one null vector of $b_{01}$ is
		needed to constrain them, and so
		$\mathrm{rank}(b_0)=\mathrm{rank}(b_{01})=1$. This again leads
		to a special case of the first type in Eq.~(\ref{eq:erset1}).
\end{enumerate}
The above exhausts all possibilities. We thus conclude that Eq.~(\ref{eq:erset1}) covers
all ERs for $n=3$. \qed

One can similarly construct the ERs for higher $n$, but it can have more possibilities. For example, for $n=4$, one may have ERs with $\mathrm{rank}(b_0)=2$.
This is possible if, for example, $b_{12}$ and $b_{34}$ vanish altogether.
After taking into account $\left(y_{kl}Q_{y,kl}+z_{kl}Q_{z,kl}\right)v^{(kl)}=0$,
we are left with 8 free parameters: $x_1,\dots x_4$, $y_{13}$, $y_{14}$,
$y_{23}$, $y_{24}$, and thus two null vectors from $v^{(0)}$ are sufficient to
constrain them.

We now derive the corresponding bounds using the ERs $Q_\mathrm{ex1}(x)$ and $Q_\mathrm{ex2}^{ij}$, and remove the
$x$-dependence that appears in the ERs of the first type.  These bounds would
be the exact ones for $n=3$, but are only conservative for $n>3$.  For the
first class of ERs, $b_0$ can be written as $xx^T$, so the bounds are
\begin{flalign}
&\mcQ\cdot\mcM
=x^T\cdot A\cdot x+|x|^T\cdot B\cdot |x| \ge0 ~ ~ \forall x\in \mb R^{n}
\nonumber\\&
A=2
\begin{bmatrix}
	2C_{1111} & C'_{1122} &\cdots & C'_{11nn}
	\\
	C'_{1122} & 2C_{2222} & \cdots & C'_{22nn}
	\\
	\vdots & \multicolumn{2}{c}{ \ddots} & \vdots
	\\
	C'_{11nn} & C'_{22nn} &\cdots & 2C_{nnnn}
\end{bmatrix},\qquad
 B=
\begin{bmatrix}
	0 & C_{1212} &\cdots & C_{1n1n}
	\\
	C_{1212} & 0 & \cdots & C_{2n2n}
	\\
	\vdots & \multicolumn{2}{c}{ \ddots} & \vdots
	\\
	C_{1n1n} & C_{2n2n} &\cdots & 0
\end{bmatrix},
\end{flalign}
where $|x|\equiv \begin{bmatrix}|x_1|&|x_2|&\dots&|x_n|\end{bmatrix}^T$.
This means that the following $2^{n-1}$ matrices

\begin{flalign}
	C_{s_1s_2\cdots s_n}\equiv	
\begin{bmatrix}
	4C_{1111} & 2s_1s_2 C'_{1122}+C_{1212} &\cdots & 2s_1s_n C'_{11nn}+C_{1n1n}
	\\
	2s_1s_2 C'_{1122}+C_{1212} & 4C_{2222} & \cdots & 2s_2s_n C'_{22nn}+C_{2n2n}
	\\
	\vdots & \multicolumn{2}{c}{ \ddots} & \vdots
	\\
	2s_1s_n C'_{11nn}+C_{1n1n} & 2s_2s_n C'_{22nn}+C_{2n2n} &\cdots & 4C_{nnnn}
\end{bmatrix}
,~ ~ s_1=1; s_2,s_3,\dots s_n=\pm1
\end{flalign}
are all PSD in the first orthant, i.e.,
\begin{flalign}
x^T\cdot C_{s_1s_2\cdots s_n}\cdot x\ge0\ \forall
x\in\mb R_+^n
\end{flalign}
For example, when $n=3$, $C_{+++}$ is given by
\begin{flalign}
	C_{+++}=2
\begin{bmatrix}
	2C_{1111} & C_{1122}+C_{1212} & C_{1133}+C_{1313}\\
	C_{1122}+C_{1212} & 2C_{2222} & C_{2233}+C_{2323}\\
	C_{1133}+C_{1313} & C_{2233}+C_{2323} & 2C_{3333}\\
\end{bmatrix}
\end{flalign}
and the condition for $C_{+++}$ to be PSD in the first octant is:
\begin{flalign}
& C_{1111}\geq 0
~\mbox{and}~ ~
 C_{2222}\geq 0
~\mbox{and}~ ~
 C_{3333}\geq 0
~\mbox{and}~ ~
 C_{2233}+C_{2323}+2 \sqrt{C_{2222} C_{3333}}\geq 0
\nonumber\\
& ~\mbox{and} ~ ~ C_{1122}+C_{1212}+2 \sqrt{C_{1111} C_{2222}}\geq 0
~\mbox{and} ~ ~ C_{1133}+C_{1313}+2 \sqrt{C_{1111} C_{3333}}\geq 0
\nonumber\\
&\mbox{and}~ ~\Bigg[
\left(2 \left(C_{1122}+C_{1212}\right) C_{3333}-\left(C_{1133}+C_{1313}\right) \left(C_{2233}+C_{2323}\right)\right) \left(4 C_{2222} C_{3333}-\left(C_{2233}+C_{2323}\right){}^2\right)\geq 0
\\
&\mbox{or}~ ~
\left(2 \left(C_{1133}+C_{1313}\right) C_{2222}-\left(C_{1122}+C_{1212}\right) \left(C_{2233}+C_{2323}\right)\right) \left(4 C_{2222} C_{3333}-\left(C_{2233}+C_{2323}\right){}^2\right)\geq 0
\nonumber \\
&\mbox{or}~ ~
2 C_{1111}\geq \frac{2 C_{3333} \left(C_{1122}+C_{1212}\right){}^2+\left(C_{1133}+C_{1313}\right) \left(2 \left(C_{1133}+C_{1313}\right) C_{2222}-2 \left(C_{1122}+C_{1212}\right) \left(C_{2233}+C_{2323}\right)\right)}{4 C_{2222} C_{3333}-\left(C_{2233}+C_{2323}\right){}^2}
\Bigg]\nonumber
\end{flalign}
Similar results can be written down for $C_{++-}, C_{+-+}, C_{+--}$.
The second class of ERs simply gives $C_{ijij}\ge0$.\\

\ssec{ERs of a generic bi-scalar EFT.}
Now consider a generic multi-scalar model, that is without $Z_2$ symmetry. We will focus on the case of $n=2$. For $n=2$, the amplitude $\mcM$ is given by
\begin{flalign}
	\mcM=
	\begin{bmatrix}
	4C_{1111} &  C'_{1122}  & C_{1112}  & C_{1112} \\
	C'_{1122} & 4C_{2222}  & C_{1222}  & C_{1222}          \\
	C_{1112} & C_{1222}  & C_{1212}  & C'_{1122}           \\
	C_{1112}& C_{1222} &  C'_{1122}  & C_{1212}           
	\end{bmatrix}\,.
\end{flalign}
We find that $\mathrm{ext}(\mbf Q^{2^4})$ can be parameterized by three parameters:
\begin{flalign}
	\mcQ(a,b,c)=
\begin{bmatrix} 
	a^2 & ab & ac & ac \\
	ab & b^2 & bc & bc \\
	ac & bc & 2c^2-ab & ab \\
	ac & bc & ab & 2c^2-ab
\end{bmatrix},
\label{biscalarERs}
\end{flalign}
where $a,b,c$ are free real numbers satisfying $c^2\ge ab$,
which are the equivalence of the $r$ parameter in the $Z_2$ symmetric case. In the following, we will prove that these ERs are complete.

It is sufficient to show that all elements of $\mbf Q^{2^4}$ can be written as
a positive linear combination of elements in Eq.~(\ref{biscalarERs}).  With
crossing symmetry, an arbitrary element $\mcQ$ can be parameterized as a matrix
\begin{flalign}
	\raisebox{-7pt}{$\mcQ=$ }
\begin{blockarray}{lccccr}
   kl=&11 & 22 & 12 & 21 &\\
\begin{block}{l[cccc]r}
	&A & B & E & E &ij=11 \\
	&B & C & F & F & 22\\
	&E & F & D & B & 12\\
	&E & F & B & D & 21\\
\end{block}
\end{blockarray}\raisebox{-7pt}{\,.}
\end{flalign}
According to the Sylvester's criterion, $\mcQ\succeq0$ implies that
$|\mcQ|\ge0$, $D\ge0$, as well as the determinant of the bottom-right
$2\times2$ block also being positive, i.e.,~$D\ge |B|$. Defining the matrix
\begin{flalign}
	\mcM'= \begin{bmatrix} 
		A & B & E \\
		B & C & F \\
		E & F & \frac{B+D}{2} 
	\end{bmatrix}\,,
\end{flalign}
we have $|\mcM'|=2|\mcQ|/(D-B)\ge0$. Since $\mcQ\succeq0$ also implies
$ \begin{bmatrix} 
		A & B\\
		B & C
	\end{bmatrix}\succeq0 $ and $A\ge0$,
we conclude that $\mcM'\succeq0$. This allows us to write down
the decomposition of $\mcM'$ as follows
\begin{flalign}
	\mcM'=V^{T}V,\quad V=
	 \begin{bmatrix} 
		a_1 & b_1 & c_1 \\
		a_2 & b_2 & c_2 \\
		a_3 & b_3 & c_3   
	\end{bmatrix}\,,
\end{flalign}
and then $\mcQ$ can be written as the sum of the following ERs
\begin{flalign}
E_i=\begin{bmatrix} 
	a_i^2 & a_ib_i & a_ic_i & a_ic_i \\
	a_ib_i & b_i^2 & b_ic_i & b_ic_i \\
	a_ic_i & b_ic_i & 2c_i^2-a_ib_i & a_ib_i \\
	a_ic_i & b_ic_i & a_ib_i & 2c_i^2-a_ib_i
\end{bmatrix},\quad i=1,2,3 ,
\end{flalign}
provided $c_i^2\ge a_ib_i$ for all $i$. To show this is possible, defining
$\Delta_i=c_i^2-a_ib_i$, we have $\sum_i\Delta_i=(D-B)/2\ge0$. 
Note that the Cholesky decomposition allows us to set $a_2=a_3=c_3=0$,
so that $\Delta_3=0$ and $\Delta_2=c_2^2\ge0$. If $\Delta_1$ is also
non-negative, then we have $\Delta_i\ge0$ for all $i$. Otherwise,
one can always replace $a_i,b_i,c_i$ by $a'_i,b'_i,c'_i$ for $i=1,2$,
the latter being defined as
\begin{flalign}
\begin{bmatrix} 
	a'_1 \\
	a'_2 
\end{bmatrix}\equiv
\begin{bmatrix} 
        \cos\theta & \sin\theta \\
       -\sin\theta & \cos\theta
\end{bmatrix}
\begin{bmatrix} 
	a_1 \\
	a_2 
\end{bmatrix}  ,
\end{flalign}
and similar for $b'_i,c'_i$. 
This changes $E_1$ and $E_2$ but will leave the sum of $E_1$, $E_2$ and $E_3$
invariant.
Now $\Delta'_1={c'_1}^2-a'_1b'_1$ is a function of $\theta$,
and we have
\begin{flalign}
	& \Delta'_1<0,\ \mbox{for }\theta=0 , \\
	& \Delta'_1=c_2^2\ge0 ,\ \mbox{for }\theta=\frac{\pi}{2}  .
\end{flalign}
The latter condition is valid because we have set $a_2=0$. Therefore we can find a
$\theta$ such that $\Delta'_1=0$ and $\Delta'_2=(D-B)/2\ge0$.
With this $\theta$, $\mcQ$ can be written as a sum of $E_1$, $E_2$ and $E_3$,
which are extremal in $\mbf Q^{2^4}$. \qed

The full positivity bounds are then given by $\mc Q(a,b,c)\cdot \mc M\ge0$.
Further removing the $a,b,c$ dependence is possible thanks to a
theorem proved by Hilbert about PSD quartic forms [68], which will be presented shortly. Before that, we want to point out that for $n=2$, all the ERs of $\mc{M}$ are just ERs of elastic amplitudes between $u_i\ket{i}$ and
$v_i\ket{i}$:
$u^iv^ju^kv^l\mcM^{ijkl}$. That is, for $n=2$, $\mcQ_{uv}^{ijkl}\equiv u^iv^ju^kv^l+v^iu^jv^ku^l$ covers the whole $\mbf Q^{n^4}$. To see this, simply take $u=(a,c-\sqrt{c^2-ab})$ and $v=(a,c+\sqrt{c^2-ab})$, and we find that $\mc Q_{uv}$ is then proportional to $\mc Q(a,b,c)$ of Eq.~(\ref{biscalarERs}).

Now, we derive the explicit expressions for positivity bounds on
the generic bi-scalar model.  First, note that the $\mc Q(a,b,c)\cdot \mc
M\ge0$ can be re-written as
\begin{flalign}
\begin{bmatrix}
	a & c & b
\end{bmatrix}
\cdot
\begin{bmatrix} 
	2C_{1111} & C_{1112} & C_{1122}   \\
	C_{1112} & 2C_{1212} & C_{1222}   \\
        C_{1122} & C_{1222} & 2C_{2222}   
\end{bmatrix}
\cdot
\begin{bmatrix}
	a \\ c \\ b
\end{bmatrix}\ge0
\,,
\label{eq:Dpsd}
\end{flalign}
for all $c^2\ge ab$.  This can be equivalently cast as
\begin{flalign}
f(r,s,w)&\equiv\begin{bmatrix}
	w^2 & \frac{rw+sw}{2} & rs
\end{bmatrix}\cdot 
\begin{bmatrix} 
	2C_{1111} & C_{1112} & C_{1122}   \\
	C_{1112} & 2C_{1212} & C_{1222}   \\
        C_{1122} & C_{1222} & 2C_{2222}   
\end{bmatrix}
\cdot
\begin{bmatrix}
	w^2 & \frac{rw+sw}{2} & rs
\end{bmatrix}^T\nonumber\\
	&\ge0 ~~~ \forall r,s,w\in\mb R,
	\label{eq:fuvw}
\end{flalign}
that is, the quartic form $f$ must be PSD.  It was proved by Hilbert that a PSD
quartic form is a sum of squares, iff it has no more than 3 variables
[56].  In our current case, $f$ only has 3 variables ($r,w,s$), so whether it is PSD can be
determined by whether one can write it as a sum of complete squares.  Note that $r^2$ and $s^2$ cannot appear
within the squares, as $f(r,s,w)$ is at most quadratic in $r,s$ respectively.
Let $f=\sum_\alpha\left( x_\alpha\cdot
\begin{bmatrix}w^2&rs&rw&sw\end{bmatrix}\right)^2=\sum_{i,j}X_{ij}W_{ij}$,
where $x_\alpha= \begin{bmatrix}x_\alpha^1&x_\alpha^2&
	x_\alpha^3&x_\alpha^4\end{bmatrix}$ and
\begin{flalign}
	W=\begin{bmatrix}	
		w^4   & rsw^2  & rw^3   & sw^3    \\
		rsw^2 & r^2s^2 & r^2sw  & rs^2w   \\
		rw^3  & r^2sw  & r^2w^2 & rsw^2   \\
		sw^3  & rs^2w  & rsw^2  & s^2w^2  \\
	\end{bmatrix}
\end{flalign}
and $X=\sum_\alpha x_\alpha x_\alpha^T \in \bfS_+^{4\times 4}$.
Comparing with Eq.~(\ref{eq:fuvw}), $X$ can be written as
\begin{flalign}
	X=\frac{1}{2}
\mcM_\mathrm{scalar}+ d
	\begin{bmatrix}	
		0&1&0&0\\
		1&0&0&0\\
		0&0&0&-1\\
		0&0&-1&0\\
	\end{bmatrix}
	\label{eq:Xd}
\end{flalign}
where $2d=\sum_\alpha \left(x_\alpha^1x_\alpha^2-x_\alpha^3x_\alpha^4\right)$ is
undetermined.  The bounds can be derived by the existence of a real parameter
$d$ such that $X$ is a $4\times4$ PSD matrix. For the non-degenerate case without any bound saturation,  using the Sylvester's criterion, this is equivalent to the following conditions:
\begin{flalign}
&
	4C_{1111}>0,\quad
	\left|
	\begin{array}{cccc}
	4C_{1111} &  C'_{1122}+2d  \\
	C'_{1122}+2d & 4C_{2222}     
	\end{array}
	\right|>0,\quad
	\left|
	\begin{array}{cccc}
	4C_{1111} &  C'_{1122}+2d  & C_{1112} \\
	C'_{1122}+2d & 4C_{2222}  & C_{1222}          \\
	C_{1112} & C_{1222}  & C_{1212}          
	\end{array}
	\right|>0
\\&
	\left|
	\begin{array}{cccc}
	4C_{1111} &  C'_{1122}+2d  & C_{1112}  & C_{1112} \\
	C'_{1122}+2d & 4C_{2222}  & C_{1222}  & C_{1222}          \\
	C_{1112} & C_{1222}  & C_{1212}  & C'_{1122}-2d           \\
	C_{1112}& C_{1222} &  C'_{1122}-2d  & C_{1212}           
	\end{array}
	\right|>0
\end{flalign}
These are polynomial inequalities for $d$, and one has to find the
constraints on coefficients such that at least one solution for $d$ exists. This can be carried out analytically, which leads to the explicit positivity bounds: 
\begin{flalign}
& C_{1111}> 0 \quad \mbox{and}\quad 4 C_{1111} C_{1212}-C_{1112}^2> 0
\nonumber\\&\mbox{and}\quad \Big\{
C_{1112} C_{1122} C_{1222}-C_{1111}
   C_{1222}^2-C_{1112}^2 C_{2222}+C_{1212}
   \left(-C_{1122}^2+4 C_{1111}
   C_{2222}\right)> 0
\nonumber\\&
\mbox{or}\quad \Big[
	\Delta\equiv
	3 \left(4 C_{1111} C_{2222}-C_{1112}
   C_{1222}\right)+\left(C_{1122}+C_{1212}\right){}^2>0
\\ & \mbox{and}\quad 
\frac{3 C_{1112}^2}{4
   C_{1111}}-2 \left(C_{1122}+C_{1212}\right)<
\sqrt{\Delta }< C_{1212}-2 C_{1122}
\nonumber\\& \mbox{and}\quad
2 \Delta ^{3/2}> 
27 \left(C_{1111} C_{1222}^2+C_{1112}^2
   C_{2222}\right)-9
   \left(C_{1122}+C_{1212}\right) \left(8
   C_{1111} C_{2222}+C_{1112} C_{1222}\right)+2
   \left(C_{1122}+C_{1212}\right){}^3 \Big]\Big\}\nonumber
\end{flalign}

\subsection{More details about SM gauge bosons}
\label{app:gauge}
Here we provide more details about the positivity bounds on the SMEFT anomalous gluon couplings.
We first list the relevant gluon operators in the $gg\to gg$ scatterings in the SMEFT.
The $P$-conserving dim-8 operators are as follows [57]
\begin{flalign}
	\begin{aligned}	
&Q^{(1)}_{G^4}=\left( G^A_{\mu\nu}G^{A\mu\nu} \right)
		\left( G^B_{\rho\sigma}G^{B\rho\sigma} \right)
\\
&Q^{(2)}_{G^4}=\left( G^A_{\mu\nu}\tilde G^{A\mu\nu} \right)
		\left( G^B_{\rho\sigma}\tilde G^{B\rho\sigma} \right)
\\
&Q^{(3)}_{G^4}=\left( G^A_{\mu\nu} G^{B\mu\nu} \right)
		\left( G^A_{\rho\sigma} G^{B\rho\sigma} \right)
	\end{aligned}
	\hspace{2cm}
	\begin{aligned}	
&Q^{(4)}_{G^4}=\left( G^A_{\mu\nu}\tilde G^{B\mu\nu} \right)
		\left( G^A_{\rho\sigma}\tilde G^{B\rho\sigma} \right)
\\
&Q^{(7)}_{G^4}=d^{ABE}d^{CDE}\left( G^A_{\mu\nu} G^{B\mu\nu} \right)
		\left( G^C_{\rho\sigma} G^{D\rho\sigma} \right)
\\
&Q^{(8)}_{G^4}=d^{ABE}d^{CDE}\left( G^A_{\mu\nu} \tilde G^{B\mu\nu} \right)
		\left( G^C_{\rho\sigma} \tilde G^{D\rho\sigma} \right)
	\end{aligned}
\end{flalign}
while the dim-6 operator
$O_{G}=f^{ABC}G^{A\nu}_{\mu}G^{A\rho}_{\nu}G^{A\mu}_{\rho}$ also enters through
double insertion diagrams. Thus, we define the coefficient vector
\begin{flalign}
\vec c\equiv\begin{bmatrix}
	C_{G^4}^{(1)} & C_{G^4}^{(2)} & C_{G^4}^{(3)} & C_{G^4}^{(4)} & C_{G^4}^{(7)} & C_{G^4}^{(8)} & c_G^2
\end{bmatrix}\,.
	\label{eq:gluonCs}
\end{flalign}
As mentioned in the main text, the optimal positivity bounds can be efficiently obtained by the SDP approach numerically, which can then be uplifted to bounds consisting of integers. The results agree with those from the symmetric extremal approach. 
The full set of 45 bounds can be presented in the form of
$\vec x\cdot \vec c\ge0$.
The $\vec x$ vectors are
\begin{flalign}
\begin{aligned}
&[0,0,0,1,0,0,0]\\ 
&[0,0,1,1,1,0,0]\\ 
&[2,0,1,0,0,0,0]\\ 
&[0,2,0,1,0,0,0]\\ 
&[0,0,3,0,2,0,0]\\ 
&[0,0,0,3,0,2,0]\\ 
&[0,0,0,0,0,0,1]\\  
&[6,0,3,0,2,0,0]\\ 
&[4,2,2,1,2,0,0]\\ 
&[0,0,4,0,0,0,-9]\\ 
&[6,0,6,0,5,0,0]\\ 
\end{aligned}\quad
\begin{aligned}
&[0,0,3,6,5,4,0]\\ 
&[0,0,6,3,7,2,0]\\ 
&[8,6,1,6,0,2,0]\\ 
&[0,6,3,12,5,0,0]\\ 
&[8,6,1,12,0,0,0]\\ 
&[0,6,6,9,10,4,0]\\ 
&[0,12,0,14,0,0,-9]\\ 
&[0,0,8,8,0,8,-27]\\ 
&[12,0,14,0,0,0,-27]\\ 
&[6,8,12,1,0,0,-27]\\ 
&[8,16,4,8,0,8,-27]\\ 
\end{aligned}\quad
\begin{aligned}
&[0,24,0,12,0,8,-27]\\ 
&[8,22,1,14,0,10,-27]\\ 
&[24,0,12,21,15,14,0]\\ 
&[24,32,24,4,8,0,-27]\\ 
&[48,36,21,27,25,0,0]\\ 
&[32,40,4,80,0,0,-27]\\ 
&[0,48,0,48,0,40,-81]\\ 
&[24,0,36,24,16,40,-81]\\ 
&[0,0,48,24,32,40,-81]\\ 
&[0,0,24,48,16,56,-81]\\ 
&[88,32,56,4,40,0,-27]\\ 
\end{aligned}\quad
\begin{aligned} 
&[96,42,27,84,25,0,0]\\ 
&[96,66,42,39,50,4,0]\\ 
&[120,42,39,42,40,14,0]\\ 
&[40,32,80,4,0,0,-189]\\ 
&[0,0,24,120,40,104,-81]\\ 
&[0,0,120,24,104,40,-81]\\  
&[48,0,96,24,0,40,-243]\\ 
&[0,192,168,96,112,120,-405]\\ 
&[168,480,168,156,56,160,-729]\\ 
&[264,384,156,168,16,200,-729]\\ 
&[288,384,216,168,0,200,-891]\\ 
&[336,768,672,216,0,200,-2187]\\
\end{aligned}
\end{flalign}

The constraining power of the new approach compared to the previous elastic method can be assessed by computing the solid angle of the
positivity region in the Wilson coefficient space.  This is done by uniformly
sampling a unit 5-sphere in the space spanned by $\{C_{G^4}^{(i)}\}$, or a
6-semisphere if $c_G^2\ge0$ is also included, and counting the fraction of the parameter space that
satisfies all positivity bounds. In Table~\ref{tab:su3}, we show a comparison between the
full bounds and the elastic bounds, the latter being computed with the ODE method
described in [30].  

\begin{table}
\begin{tabular}{|c|c|c|}
\hline & With $c_G^2\ge0$ & $c_G=0$  \\
\hline Full bounds & $1.6628\% \pm 0.0007\%$ & $6.3349\% \pm 0.0012\%$  \\
\hline Elastic bounds & $1.818\% \pm 0.007\%$ & $6.584\% \pm 0.013\%$  \\
\hline
\end{tabular}
\caption{Comparison between the volume of the positive region of the exact
positivity bounds and that of the elastic bounds, for 4-gluon SMEFT operators.}
\label{tab:su3}
\end{table}

\subsection{More details about SM fermions}
\label{app:flavor}

Here we provide more details about the positivity bounds on the SMEFT fermion operators.
First, we derive the fermion scattering amplitude
in a crossing symmetric form.
The forward limit amplitude for $f_if_j\to f_kf_l$ with $f_i=(e_R,\mu_R,\tau_R)$
is not crossing symmetric, as $f_if_j\to f_kf_l$ is related to, for example,
$f_i\bar f_l\to f_k\bar f_j$ under $(j\leftrightarrow l)$ crossing.
However, if we consider a larger multiplet by combining $f_1,f_2,f_3$ with 
$\bar f_1,\bar f_2,\bar f_3$, then the forward amplitude would have 4
diagonal blocks, which correspond to $f_if_j\to f_kf_l$, $f_i\bar f_j\to
f_k\bar f_l$, $\bar f_if_j\to \bar f_kf_l$ and $\bar f_i\bar f_j\to \bar
f_k\bar f_l$ respectively.  This is because $f\bar f\to \bar f f$
vanishes because of angular momentum conservation, while $f f\to \bar f \bar f$
vanishes due to the U(1) hypercharge carried by $f$. 

Denoting the four diagonal blocks by $\mcM_{1}$, $\mcM_{2}$, $\mcM_{3}$,
$\mcM_{4}$.  Assuming CP-conservation, $C_{ijkl}=C_{jilk}$ are real, and we have
$\mcM_{1}=\mcM_{4}$ and
$\mcM_{2}=\mcM_{3}$.  If we further take the particle basis to be $f^R_i=(f_i+\bar
f_i)/2$ and $f^I_i=(f_i-\bar f_i)/(2i)$, the full amplitude is crossing symmetric:
\begin{flalign}
	\raisebox{-7pt}{$\mcM=$ }
\begin{blockarray}{ccccc}
   f^Rf^R & f^If^I & f^Rf^I & f^If^R &\\
\begin{block}{[cccc]l}
   \mcM_A & \mcM_B & 0 & 0  & f^Rf^R \\
   \mcM_B & \mcM_A & 0 & 0  & f^If^I \\
   0 & 0 & \mcM_A & -\mcM_B & f^Rf^I \\
   0 & 0 & -\mcM_B & \mcM_A & f^If^R \\
\end{block}
\end{blockarray}
\end{flalign}
where $\mcM_A=(\mcM_1+\mcM_2)/2$ and $\mcM_B=(-\mcM_1+\mcM_2)/2$ are $9\times9$ matrices,
whose rows and columns correspond to $(i,j)$ and $(k,l)$ indices respectively,
both in the sequence of $(1,1),(1,2),(1,3),(2,1),\dots,(3,3)$.
We have $\mcM\in
\overrightarrow\bfS^{6^4}$. Therefore the bounds of $\mcM$ 
is dual to $\mbf Q^{6^4}$ and can be solved
using the new SDP method. 
The explicit expressions for $\mcM_A$ and $\mcM_B$ are given below
\begin{flalign}
	&\mcM_A=\begin{bmatrix}
		M^{(A11)}_\mathbf{3\times3} & M^{(A12)}_\mathbf{3\times3} &
		M^{(A13)}_\mathbf{3\times3}\\
		M^{(A12)}_\mathbf{3\times3} & M^{(A22)}_\mathbf{3\times3} &
		M^{(A23)}_\mathbf{3\times3}\\
		M^{(A13)}_\mathbf{3\times3} & M^{(A23)}_\mathbf{3\times3} &
		M^{(A33)}_\mathbf{3\times3}\\
	\end{bmatrix},\qquad
	\mcM_B=\begin{bmatrix}
		0_\mathbf{3\times3} & M^{(B12)}_\mathbf{3\times3} &
		M^{(B13)}_\mathbf{3\times3}\\
		-M^{(B12)}_\mathbf{3\times3} & 0_\mathbf{3\times3} &
		M^{(B23)}_\mathbf{3\times3}\\
		-M^{(B13)}_\mathbf{3\times3} & -M^{(B23)}_\mathbf{3\times3} &
		0_\mathbf{3\times3}\\
	\end{bmatrix}  ,
\end{flalign}
where the $M^{(Aij)}$ and $M^{(Bij)}$ matrices are symmetric and antisymmetric
respectively, so $\mcM_A$ and $\mcM_B$ are both symmetric.  In particular,
\begin{flalign}
	&\mcM_A^{ijkl}=\mcM_A^{ilkj}=\mcM_A^{kjil}=\mcM_A^{jilk}  ,
	\\
	&\mcM_B^{ijkl}=-\mcM_B^{ilkj}=-\mcM_B^{kjil}=\mcM_B^{jilk}  ,
\end{flalign}
and $M^{(Aij)}$ and $M^{(Bij)}$ are given by:
\begin{flalign}
&M^{(A11)}= -2\begin{bmatrix}
 2 C_{1111} & C_{1112} & C_{1113} \\
 C_{1112} & C_{1221} & C_{1231} \\
 C_{1113} & C_{1231} & C_{1331} \\
\end{bmatrix},~
M^{(A22)}=
-2\begin{bmatrix}
 C_{1221} & C_{1222} & C_{1223} \\
 C_{1222} & 2 C_{2222} & C_{2223} \\
 C_{1223} & C_{2223} & C_{2332} \\
\end{bmatrix},~
M^{(A33)}=
-2\begin{bmatrix}
 C_{1331} & C_{1332} & C_{1333} \\
 C_{1332} & C_{2332} & C_{2333} \\
 C_{1333} & C_{2333} & 2 C_{3333} \\
\end{bmatrix}
\end{flalign}
\begin{flalign}
&M^{(A12)}=-
\begin{bmatrix}
2 C_{1112} & C_{1122}+2 C_{1212} & C_{1123}+C_{1213} \\
 C_{1122}+2 C_{1212} & 2 C_{1222} & C_{1232}+C_{1322} \\
 C_{1123}+C_{1213} & C_{1232}+C_{1322} & 2 C_{1332} \\
\end{bmatrix}
\nonumber\\
&M^{(A13)}=-
\begin{bmatrix}
 2 C_{1113} & C_{1123}+C_{1213} & C_{1133}+2 C_{1313} \\
 C_{1123}+C_{1213} & 2 C_{1223} & C_{1233}+C_{1323} \\
 C_{1133}+2 C_{1313} & C_{1233}+C_{1323} & 2 C_{1333} \\
\end{bmatrix}
\nonumber\\
&M^{(A23)}=-
\begin{bmatrix}
2 C_{1231} & C_{1232}+C_{1322} & C_{1233}+C_{1323} \\
 C_{1232}+C_{1322} & 2 C_{2223} & C_{2233}+2 C_{2323} \\
 C_{1233}+C_{1323} & C_{2233}+2 C_{2323} & 2 C_{2333} \\
\end{bmatrix}
\end{flalign}

\begin{flalign}
&M^{(B12)}=
\begin{bmatrix}
 0 & 2 C_{1212}-C_{1122} & C_{1213}-C_{1123} \\
 C_{1122}-2 C_{1212} & 0 & C_{1322}-C_{1232} \\
 C_{1123}-C_{1213} & C_{1232}-C_{1322} & 0 \\
\end{bmatrix}
\nonumber\\
&M^{(B13)}=
\begin{bmatrix}
 0 & C_{1213}-C_{1123} & 2 C_{1313}-C_{1133} \\
 C_{1123}-C_{1213} & 0 & C_{1323}-C_{1233} \\
 C_{1133}-2 C_{1313} & C_{1233}-C_{1323} & 0 \\
\end{bmatrix}
\nonumber\\
&M^{(B23)}=
\begin{bmatrix}
 0 & C_{1322}-C_{1232} & C_{1323}-C_{1233} \\
 C_{1232}-C_{1322} & 0 & 2 C_{2323}-C_{2233} \\
 C_{1233}-C_{1323} & C_{2233}-2 C_{2323} & 0 \\
\end{bmatrix}  .
\end{flalign}

Finally, for illustration purposes, we take a further simplification
by imposing the condition $\mcM_B=\mathbf{0}_{9\times9}$, so that $\mcM_A$ can
be rewritten as:
\begin{flalign}
&M^{(A12)}=-2
\begin{bmatrix}
 C_{1112} & 2 C_{1212} & C_{1213} \\
 2 C_{1212} & C_{1222} & C_{1232} \\
 C_{1213} & C_{1232} & C_{1332} \\
\end{bmatrix},~
M^{(A13)}=-2
\begin{bmatrix}
 C_{1113} & C_{1213} & 2 C_{1313} \\
 C_{1213} & C_{1223} & C_{1323} \\
 2 C_{1313} & C_{1323} & C_{1333} \\
\end{bmatrix},~
M^{(A23)}=-2
\begin{bmatrix}
 C_{1231} & C_{1232} & C_{1323} \\
 C_{1232} & C_{2223} & 2 C_{2323} \\
 C_{1323} & 2 C_{2323} & C_{2333} \\
\end{bmatrix}
\end{flalign}
while $M^{(A11)}$, $M^{(A22)}$ and $M^{(A33)}$ remain unchanged.
Under this restriction, the non-vanishing amplitudes are fully captured by
$\mcM_A\in\mbf C^{3^4}$, which depends on 21 independent Wilson
coefficients, and can be treated in the same way as in a three-scalar EFT. 
The bounds are given by ext($\mbf Q^{3^4}$).
The crossing symmetric amplitude $\mcM$ in the main text refers to this
$\mcM_A$.

\begin{figure}[htb]
	\begin{center}
		\includegraphics[width=.4\linewidth]{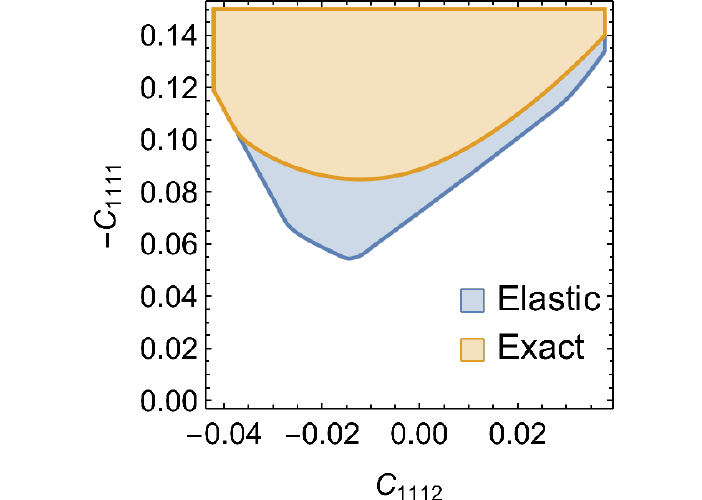}
	\end{center}
	\caption{Bounds on the flavor-conserving
coefficient $|C_{1111}|$ as a function of $C_{1112}$, with other coefficients
fixed at $\vec C_0$. We see that flavor-conserving signal of new physics is
bounded by the flavor-violating ones from below, and our 
approach improves the elastic bounds.}
	\label{fig:flavor2}
\end{figure}
To further illustrate the improvement of our new approach, in Figure~\ref{fig:flavor2}, we show the lower bounds on the flavor-conserving
coefficient $|C_{1111}|$ as a function of $C_{1112}$, with other coefficients
fixed at $\vec C_0$. This plot indicates that flavor-conserving signal
of new physics (such as $e^+e^-$ scattering)
is bounded by the flavor-violating ones from below (such as $\mu\to3e$), and
our new approach is crucial for fully capturing this information.

\subsection{Applications in constraining spin-2 EFTs}

Here we present another interesting application of our formalism for spin-2 EFTs.
General relativity, when viewed as an EFT on Minkowski space, is a theory with a massless spin-2 particle, the graviton. The discovery of the late time cosmic acceleration has prompted the proposal that the graviton may have a finite, cosmological scale Compton wavelength, i.e., a small, Hubble scale mass. The theory of massive gravity has traditionally been regarded as ill-defined due to theoretical inconsistencies such as the ghost instabilities until the dRGT model [67] was discovered which eliminates the ghost degree of freedom and is unique up to 2 free parameters $\kappa_3$ and $\kappa_4$ (or $c_3$ and $d_5$ with the relations $\kappa_{3}=2-4 c_{3}, ~\kappa_{4}=1-4 c_{3}-8 d_{5}$). The dRGT model consists of the standard Einstein-Hilbert term plus a nonlinear potential for the graviton, and can be compactly written down with the help of the square root of the metric, but for our purposes the leading Lagrangian is simply given by
\be
 \mathcal{L}_{\mathrm{dRGT}}=\frac{M_{P}^{2}}{2} \sqrt{-g} R+\frac{m^{2}}{4}\left[\epsilon \epsilon I I h h+\frac{\kappa_{3}}{M_{P}} \epsilon \epsilon I h h h+\frac{\kappa_{4}}{M_{P}} \epsilon \epsilon h h h h+...\right] ,
\ee
where $R$ is the Ricci scalar, $M_{P}$ is the Planck mass, $m$ is the graviton mass, $h_{\mu\nu}=g_{\mu\nu}-\eta_{\mu\nu}$ and $\epsilon \epsilon I h h h=-\epsilon_{\mu \nu \rho \sigma} \epsilon^{\alpha \beta \gamma \delta} \delta_{\alpha}^{\mu} h_{\beta}^{\nu} h_{\gamma}^{\rho} h_{\delta}^{\sigma}$ and so on. There has been suggestions that the dRGT model might not have an analytical UV completion (See, for example, [69]),
thus violating positivity bounds. However, Ref.~[30] has applied elastic positivity bounds to the dRGT model and found that $c_3$ and $d_5$ are compatible with positivity bounds within a finite island in the parameter space. A massive graviton has 5 independent polarization modes, and the exact bounds can be obtained by a SDP on $\mbf Q^{5^4}$. 
 We find a small improvement in the left bottom
corner of the 2D finite region, improving the minimum $d_5$ value; see the left plot of Figure
\ref{fig:dRGT}.  This suggests that dRGT theory is robust even beyond
elastic positivity.

\begin{figure}[tb]
	\begin{center}
		\includegraphics[width=.65\linewidth]{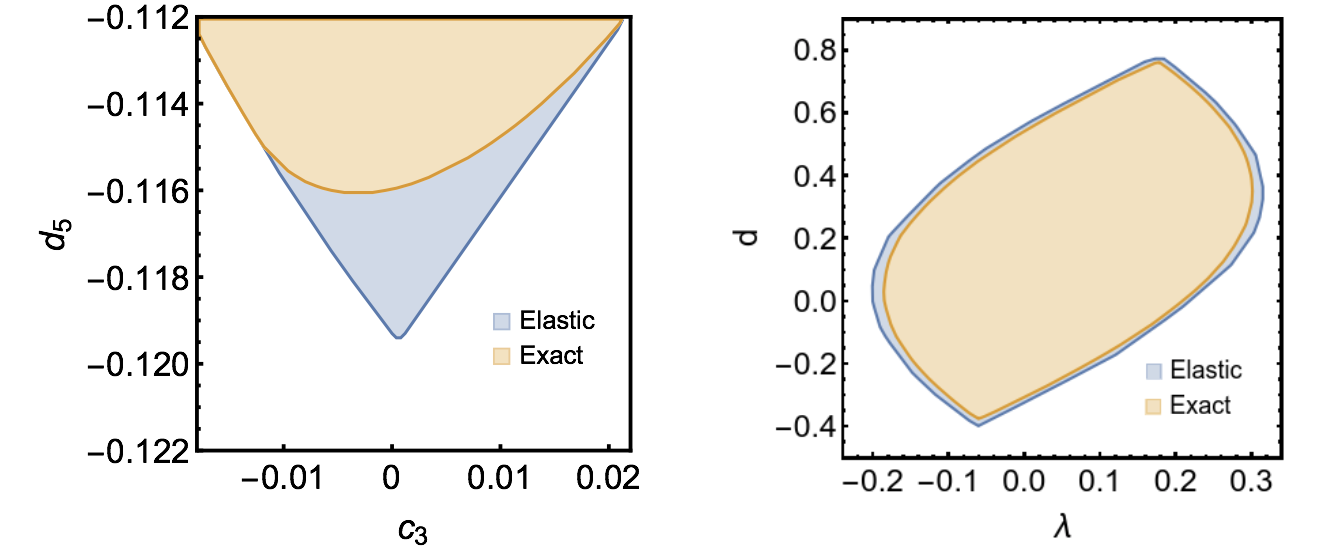}
	\end{center}
	\caption{Improvement in constraining the dRGT parameters $c_3$ and
	$d_5$ (left) and the $Z_2$ bi-field spin-2 EFT
	parameters $d$ and $\lambda$ (right; see Ref.~[66]).
	``Elastic'' denotes the superposed elastic positivity bounds, while
``Exact'' bounds are obtained by SDP. }
	\label{fig:dRGT}
\end{figure}

Elastic positivity has also been used to constrain
multi-field spin-2 EFTs [38,\,27], which are relevant in various contexts (see [66] and reference
therein). With more modes, the improvements from our approach can become more
prominent. In the right plot of Figure \ref{fig:dRGT}, we have shown an exemplary comparison between the full bounds and the elastic bounds for the $Z_2$ bi-field cycle theory [66]:
\be
\label{bispin2}
\begin{aligned}
g_{*}^{2} \mathcal{L}_{\text {cycle }}=& \frac{M_{P}^{2}}{2} \sqrt{-g^{(1)}} R\left(g^{(1)}\right)+\frac{m^{2}}{4}\left[\epsilon \epsilon I I h h+\frac{\kappa_{3}}{M_{P}} \epsilon \epsilon I h h h+\frac{\kappa_{4}}{M_{P}^{2}} \epsilon \epsilon h h h h+...\right] \\
&+\frac{M_{P}^{2}}{2} \sqrt{-g^{(2)}} R\left(g^{(2)}\right)+\frac{m^{2}}{4}\left[\epsilon \epsilon I I f f+\frac{\kappa_{3}}{M_{P}} \epsilon \epsilon I f f f+\frac{\kappa_{4}}{M_{P}^{2}} \epsilon \epsilon f f f f+...\right] \\
&+\frac{m^{2} c}{2 M_{P}} \epsilon \epsilon I h h f+\frac{m_{2}^{2} c}{2 M_{P}} \epsilon \epsilon I f f h+\frac{m^{2} \lambda}{2 M_{P}^2} \epsilon \epsilon h h f f+\frac{m^{2} d}{4 M_{P}^{2}} \epsilon \epsilon h h h f+\frac{m^{2} d}{4 M^{P}} \epsilon \epsilon h f f f+\ldots ,
\end{aligned}
\ee
where now we have two spin-2 fields $h_{\mu\nu}=g^{(1)}_{\mu\nu}-\eta_{\mu\nu}$ and $f_{\mu\nu}=g^{(2)}_{\mu\nu}-\eta_{\mu\nu}$. 
The results are similar for
other cross sections of the parameter space. For example,  see Figure \ref{fig:cdspin2} for the bounds on the subspace of $c$ and $d$. Apart from improving the bounds,
evaluations of the elastic bounds with
the ODE method [30] can be extremely inefficient
for bi-field spin-2 EFTs: the results
converge slowly with the number of the random initial conditions needed to
seed the ODE evolution. In comparison, the SDP approach is far more efficient.

\begin{figure}[htb]
	\begin{center}
		\includegraphics[width=.3\linewidth]{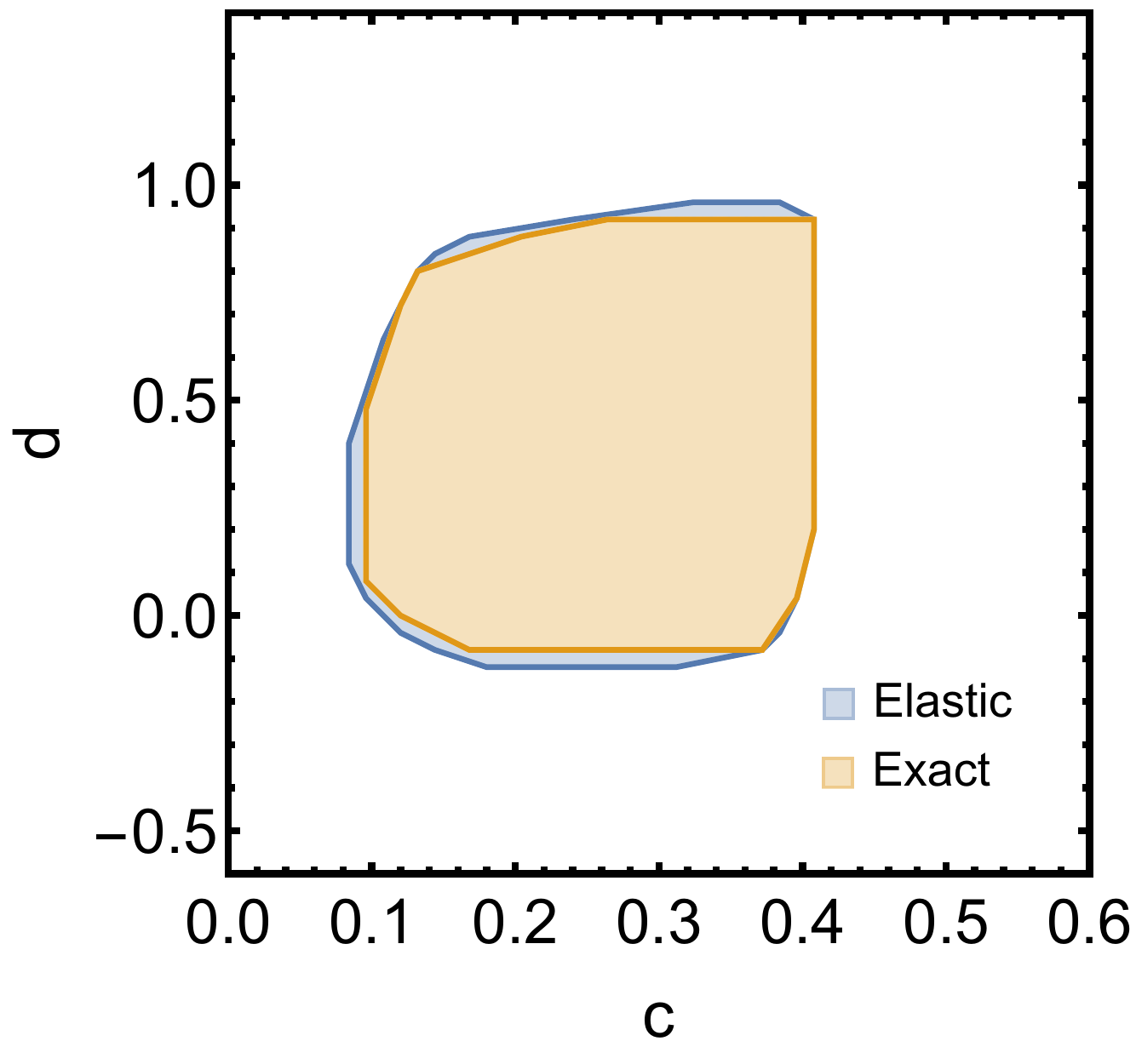}
		\includegraphics[width=.3\linewidth]{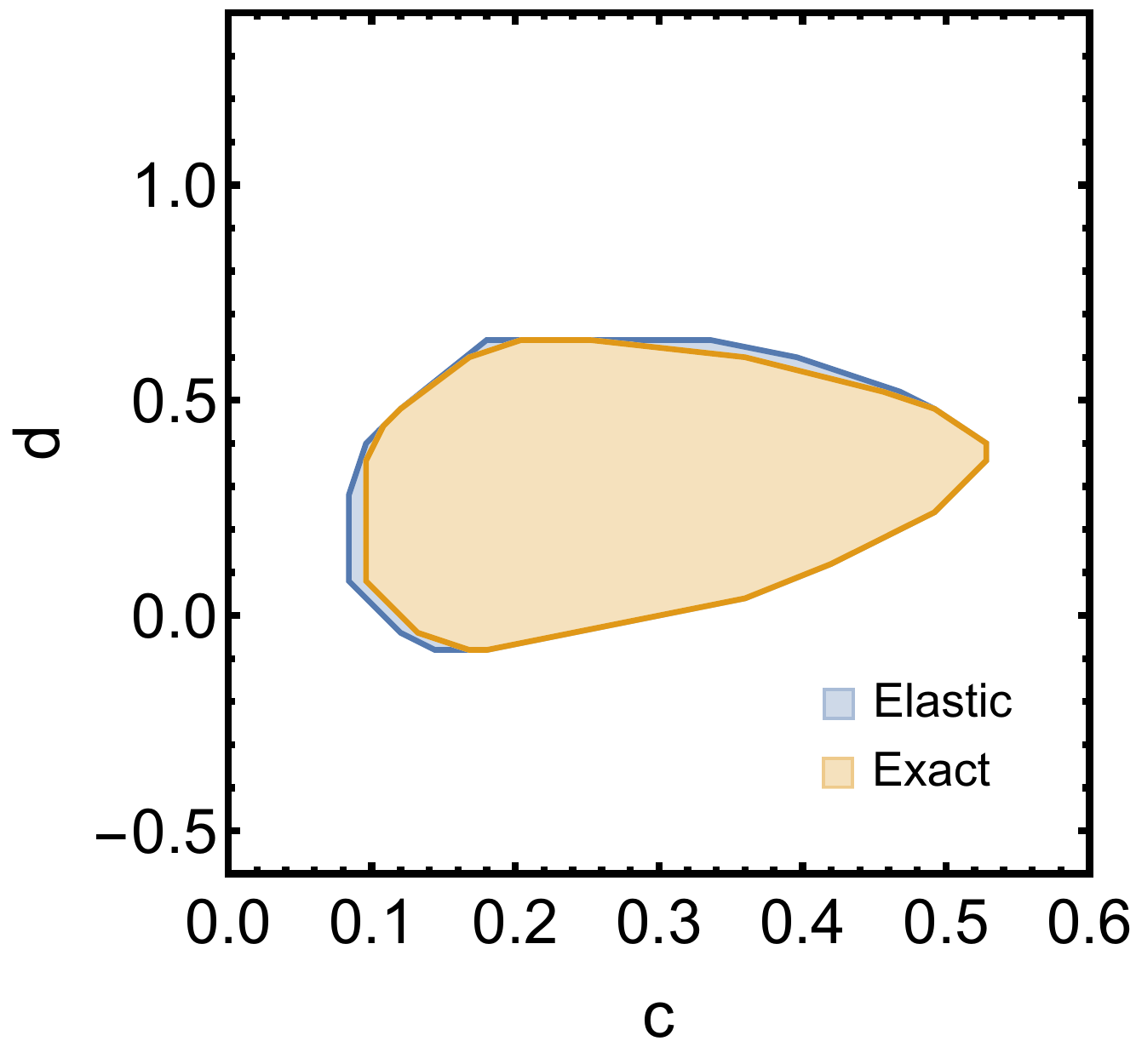}
	\end{center}
	\caption{Comparison between the elastic and full positivity bounds for $c$-$d$ cross sections in the bi-field cycle theory (\ref{bispin2}). Other parameters are chosen as $(\kappa_3,\kappa_4,\lambda)=(1.4,0.38,0.2)$ for the left figure and $(\kappa_3,\kappa_4,\lambda)=(1.1,0.38,0.2)$ for the right figure.}
	\label{fig:cdspin2}
\end{figure}

\end{widetext}

\end{document}